\documentclass[preprint,showpacs,preprintnumbers,amsmath,amssymb,nofootinbib]{revtex4}

\usepackage{etex}

\usepackage{amsthm,amscd,amsbsy,array}
\usepackage{bm}% bold math
\usepackage{soul} % underline, strikethrough, etc.

\usepackage{graphics,graphicx,xcolor}

\usepackage{soul} % underline, strikethrough,

\usepackage[colorlinks=true, pdfstartview=FitV, linkcolor=blue, citecolor=blue, urlcolor=blue]{hyperref} % hyperref
\newcommand{\cyan}{\textcolor{cyan}}
\newcommand{\magenta}{\textcolor{magenta}}
\newcommand{\red}{\textcolor{red}}
\newcommand{\blue}{\textcolor{blue}}

\newcommand{\gb}{\colorbox{green}}

\newenvironment{redtext}{\color{red}}
{\ignorespacesafterend}
\newenvironment{bluetext}{\color{blue}}{\ignorespacesafterend}

\newenvironment{magentatext}{\color{magenta}}{\ignorespacesafterend}
\newenvironment{cyantext}{\color{cyan}}{\ignorespacesafterend}

\newcommand{\bmagenta}{\begin{magentatext}}
\newcommand{\emagenta}{\end{magentatext}}
\newcommand{\bcyan}{\begin{cyantext}}
\newcommand{\ecyan}{\end{cyantext}}
\newcommand{\bblue}{\begin{bluetext}}
\newcommand{\eblue}{\end{bluetext}}
\newcommand{\bred}{\begin{redtext}}
\newcommand{\ered}{\end{redtext}}

\numberwithin{equation}{section}

\let\ssection=\section
\renewcommand{\section}{\setcounter{equation}{0}\ssection}
\textheight=%245mm
250mm %220mm %185mm
\newcommand{\cA}{{\mathcal{A}_{+}}}
\newcommand{\cB}{{\mathcal{B}_{+}}}

\newcommand{\bxi}{\boldsymbol{\xi}}
\newcommand{\bzeta}{\boldsymbol{\zeta}}

\def\aand{{\quad\text{and}\quad}}
\def\where{{\quad\text{where}\quad}}
\def\with{{\quad\text{with}\quad}}

\newcommand{\cH}{{\mathcal{H}}}

\newcommand{\bp}{{\bf p}}

\newcommand{\bx}{{\bm{x}}}

\renewcommand{\Re}{\mathrm{Re}}

\newcommand{\bw}{{\bf w}}

\newcommand{\bX}{{\bf X}}

\def\smallover#1/#2{\hbox{$\textstyle\frac{#1}{#2}$}} %

\def\bp{{\bm{p}}}

\def\parag{\hfil\break} %%%%% paragraph
\def\kikezd{\parag\underbar}

\def\benu{\begin{enumerate}}
\def\eenu{\end{enumerate}}
\def\beq{\begin{equation}}
\def\eeq{\end{equation}}
\def\beqa{\begin{eqnarray}}
\def\eeqa{\end{eqnarray}}
\def\nn{\nonumber}
\def\barray{\left(\begin{array}}
\def\earray{\end{array}\right)}
\def\barraynb{\begin{array}}
\def\earraynb{\end{array}}

\def\GWs{{gravitational waves\;}}

\def\BB{{Bialynicki-Birula and Charzynski\;}}

\def\?{\quad{\gb{\fbox{\texttt{?}}\;}}\quad}
\def\p{{\partial}}

\def\v0{\mathbf{0}}

\def\beq{\begin{equation}}
\def\eeq{\end{equation}}
\def\bea{\begin{eqnarray}}
\def\eea{\end{eqnarray}}

\def\p{\partial}

\def \p{{\partial}}

\def\cabove(#1){\stackrel{\smallcirc}{#1}}
\def\ccabove(#1){\stackrel{\smallcirc\smallcirc}{#1}}
\def\6{\partial}
\def\7{\tilde}
\def\8{\widehat}
 \def\bx{{\bf x}}

\def\G11{\Gamma_{11} }

\newcommand{\const}{\mathop{\rm const.}\nolimits}
\newcommand{\half }{\smallover{1}/{2}}

\def\smallover#1/#2{\hbox{$\textstyle\frac{#1}{#2}$}} %
\def\smallcirc{{\raise 0.5pt \hbox{$\scriptstyle\circ$}}}
\def\2{{\smallover1/2}}

\newcommand{\bigbox}[1]{\fbox{%
\rule[-20pt]{0pt}{45pt}$\;\;\displaystyle{#1}\;\;$}
}
\newcommand{\medbox}[1]{\fbox{%
\rule[-10pt]{0pt}{25pt}$\;\;\displaystyle{#1}\;\;$}%
}

\let\ssection=\section
\renewcommand{\section}{\setcounter{equation}{0}\ssection}

\def\besub{\begin{subequations}}
\def\esub{\end{subequations}}
\begin{document}

%\preprint{\texttt{2108.00838v2 [gr-qc]}}

\title{Particle motion in circularly polarized vacuum pp waves}

\author{
P. M. Zhang$^{1}$\footnote{corresponding author. mailto:zhangpm5@mail.sysu.edu.cn},
M. Elbistan$^{2}$\footnote{mailto: elbistan@itu.edu.tr},
P. A. Horvathy$^{3}$\footnote{mailto:horvathy@lmpt.univ-tours.fr}
}

\affiliation{
${}^1$ School of Physics and Astronomy, Sun Yat-sen University, Zhuhai, China
\\
${}^2$ Physics Department,
Bo\u{g}azi\c{c}i University,
34342 Bebek / Istanbul, (Turkey)
\\
${}^3$ Institut Denis Poisson CNRS/UMR 7013 - Universit\'e de Tours - Universit\'e d'Orl\'eans Parc de Grandmont, 37200; Tours, (France)
\\
%\yb{\texttt{Lukash-II-v2} }
}

\date{\today}


\begin{abstract}
Bialynicki-Birula and Charzynski argued that a gravitational wave emitted during the merger of a black hole binary may be approximated by a circularly polarized wave
 which may in turn trap particles [1]. In this paper we consider particle motion in a class of gravitational waves which includes, besides circularly polarized periodic waves (CPP) [2], also the one proposed by Lukash [3] to study anisotropic cosmological models. Both waves have a 7-parameter conformal symmetry which contains, in addition to the generic 5-parameter (broken) Carroll group, also a 6th isometry. The Lukash spacetime can be transformed by a conformal rescaling of time to a perturbed CPP problem. Bounded geodesics, found both analytically and numerically, arise when the Lukash wave is of Bianchi type VI. Their symmetries can also be derived from the Lukash-CPP relation. Particle trapping is discussed.
 \\
 \vskip2mm\noindent
 Class. Quant. Grav. \textbf{39} (2022) no.3, 035008
doi:10.1088/1361-6382/ac43d2
[arXiv:2108.00838 [gr-qc]].
\end{abstract}

\maketitle

\tableofcontents

%%%%%%%%%%%%%%%%%%%%%%%%%%%%%%%%%%%%%%%%%%%%%%%%%%%%%%%%%%%%%%%%%%%%%%%%%%%%%%
%%%%%%%%%%%%%%%%%%%%%%%%%%%%%%%%%%%%%%%%%%%%%%%%%%%%%%%%%%%%%%%%%%%%%%%%%%%%%%
\section{Introduction}\label{Intro}
%%%%%%%%%%%%%%%%%%%%%%%%%%%%%%%%%%%%%%%%%%%%%%%%%%%%%%%%%%%%%%%%%%%%%%%%%%%%%%
%%%%%%%%%%%%%%%%%%%%%%%%%%%%%%%%%%%%%%%%%%%%%%%%%%%%%%%%%%%%%%%%%%%%%%%%%%%%%%

\BB (BBC) \cite{BB} argued that the  \GWs
emitted during the merger of compact binaries
 may trap massive particles.
 Their clue is that in the vicinity of the wave axis a gravitational wave carrying angular momentum can be approximated by a Bessel beam. Then for small deviations the geodesic deviation equations yield a coupled system, their eqn. \# (14), which admits bounded solutions shown in their fig. 1 -- just like their electromagnetic counterparts do \cite{BB04,Ilderton,BBBNew}.

Their result is consistent with what was found before for Circularly Polarized Periodic (CPP) \GWs  \cite{exactsol}, whose geodesic equation can be reduced to
similar equations \cite{POLPER,IonGW}.
 Our paper extends and amplifies these findings by studying, both analytically and numerically, geodesic motion in \emph{Lukash plane gravitational waves} considered before in the study of the isotropy/anisotropy of cosmological models \cite{LukashJETP75,exactsol,Ehlers}.

%%%%%
Gravitational plane waves have a generic 5-parameter isometry group \cite{Sou73, BoPiRo,exactsol}, identified as a ``broken Carroll''  group \cite{LL,Carroll4GW}. CPP and Lukash waves are special in that they have an additional 6th ``screw''  (or ``helical'' \cite{CKlein}) isometry \cite{exactsol,POLPER,Ilderton}. The isometry group extends to a 7-parameter conformal symmetry \cite{Ehlers,Conf4pp,Conf4GW}.

The  6th isometry, which is a sort of ``spiralling time translation'', arises for our  circularly polarized waves.
  Moreover, according to Table 24.2 of  \cite{exactsol} p.385, CPP and Lukash are the only vacuum pp waves with this property.
\goodbreak

%%%%%%
The properties of CPP waves are widely known \cite{exactsol,Ehlers}; our principal interest here is to study the analogous but less-known Lukash waves.
Our clue is to reduce the  geodesic motion in a
 Lukash metric to one in a perturbed CPP wave by a clever rescaling ot  time, \eqref{lnchange} below. \emph{Exact analytic solutions} can be found by following a ``road map'' outlined in sec. \ref{SLsol}. The solvability comes from transforming the system to one with constant coefficients, \eqref{abeqmot} -- which implies also the extra 6th symmetry.
 Bounded motions arise when the Lukash metric has Bianchi type VI \footnote{
 Our earlier investigations \cite{Lukash-I} concern Lukash waves with a different range of the parameters and of different Bianchi type.}.

Our strategy fits into the  framework proposed by Gibbons \cite{GWG_Schwarzian}.
A bonus obtained from our CPP $\leftrightarrow$ Lukash correspondence is to relate their 6th isometries, see sec.\ref{IsoSect}.

The subtle notion of particle trapping is discussed in sec.\ref{bubSec}.

Before starting our study we fix our conventions. Lower-case latin letters as ${\bf x}= (x^1, x^2), u, v$ refer to generic pp waves. Latin  capitals $\bX=(X^1,X^2),\,U,V,\,Z = \frac{X^1 + iX^2}{\sqrt{2}}$  refer to Lukash waves with profile functions
${\cA}(U)$ and ${\mathcal{A}_{\times}}(U)$, respectively; greek letters
$\bxi=(\xi, \eta),\, \zeta= \frac{\xi + i\eta}{\sqrt{2}}$, completed with $T$ and $\nu$ and profile $\cB(T)$ and $\mathcal{B}_{\times}(T)$ refer to CPP(-type) waves.
%%%

\goodbreak
%%%%%%%%%%%%%%%%%%%%%%%%%%%%
\section{Circularly polarized \GWs}\label{CPGW}
%%%%%%%%%%%%%%%%%%%%%%%%%%%%%

Both CPP and Lukash waves are plane gravitational waves described, in Brinkmann coordinates, in terms of a symmetric and traceless $2\times2$ matrix
$K_{ij}(u)$ \cite{exactsol,
LukashJETP75,Ehlers,Lukash-I},
\beq
ds^2=d\bx^2 + 2 du dv + K_{ij}(u) x^i x^j du^2\,,
\label{Bmetric}
\eeq
where $\bx=\big(x^1,x^2\big)$.
The vacuum Einstein equations reduce to $\bigtriangleup\big(K_{ij}(u){x^i}{x^j}\big)=0$, where $\bigtriangleup\equiv \bigtriangleup_2$ is the transverse-space Laplacian. \eqref{Bmetric}-\eqref{Bprofile} is therefore  an exact plane wave for any profile $K_{ij}(u)$.

The profile is decomposed into $+$ and $\times$ polarization states,
\beq
K_{ij}(u){x^i}{x^j}=
\half{\cA}(u)\big((x^1)^2-(x^{2})^2\big)\,+\,\mathcal{A}_{\times}(u)\,x^1x^2\,.
\label{Bprofile}
\eeq

A spin-zero particle moves along a geodesic, described by,
%%%%%%%%%%%%%%
\begin{subequations}%\vspace{-3mm}
\begin{align}
& \bx'' - K(u)\,\bx = 0
\quad\where\quad
K(u)= \half\barray{lr}
{\cA} &{\mathcal{A}_{\times}}\\
{\mathcal{A}_{\times}} & -{\cA}\earray,
\label{ABXeq}
\\[8pt]
& v'' + {\cA}'
\left((x^1)^2 - (x^2)^2 \right )
+{\cA}\left(x^1{(x^1)}' - x^2{(x^2)}'\right )
+{\mathcal{A}_{\times}}' x^1x^2  + {\mathcal{A}_{\times}}\left(x^2{(x^1)}' + x^1{(x^2)}'\right)
= 0\,,
\label{ABVeq}
\end{align}
\label{ABeqs}
\end{subequations}
%%%%%%%%%%
where the prime means derivative w.r.t. the affine parameter $u$, $\{\,\cdot\,\}^{\prime}= d/du$\,.

For any affine parameter $\sigma$ the quantity $-g_{\mu\nu}\dot{x}^\mu \dot{x}^\mu=m^2$  where the ``dot'' means $d/d\sigma$
is a constant of the motion, identified with the relativistic mass-square.
 The transverse motion described by \eqref{ABXeq} does not depend on $m$, and the solution of  the $V$-equation  differs from that for $m=0$  by the simple shift  $m^2 u$ \cite{EDAHKZ}, allowing us to restrict our attention at \emph{massless geodesics}. After solving the transverse equations \eqref{ABXeq} the $v$-equation is solved  by,
\beq
v(u)=v_0-S(u)\,,
\label{Vlift}
\eeq
where $S(u)$ is the Hamiltonian action calculated along the transverse trajectory \cite{Bargmann,dgh91,EDAHKZ}. Therefore it is enough to solve the transverse equations \eqref{ABXeq}.

In what follows, we restrict our attention at circularly polarized  waves.
%%%%%%
A CPP wave has, for example, the profile (with a slight change of notations ${\cal A} \to {\cal B}$ and $u \to T$).
\beq
\cB(T)= - C\,\cos\big(\omega T\big),
\qquad
\mathcal{B}_{\times}(T)= C\,\sin\big(\omega T\big)\,,
\where C\,, \omega = \const
\label{CPPprofile}
\eeq
The geodesic motion in a CPP can be determined both numerically and analytically \cite{POLPER,IonGW}.

%%%%%%%%%%%%%%%%%%%%%%%%%%%%
\subsection{Lukash Geodesics}\label{LgeoSec}
%%%%%%%%%%%%%%%%%%%%%%%%%%%%

A Lukash plane gravitational wave is described, in  complex  form \cite{SiklosAll}, by
%%%%%
\beq
ds^2_L =
 2dU dV + 2 dZ d\bar{Z} - 2 C \Re[ U^{2(i\kappa -1)}Z^2] dU^2\,,
\where  Z = \frac{X^1 + iX^2}{\sqrt{2}}\,.
\label{Lmetric}
\eeq
 The coordinates are well-defined for either $U>0$ or $U<0$ but break down at $U=0$. The nature of the singularity at $U=0$ was the subject of intensive investigations  \cite{Collins,LukashJETP75,SiklosAll}.
The constant $C$  determines the strength of the wave;
$C \geq 0$ can be chosen with no loss of generality.
 $\kappa$ is the frequency (inverse wavelength) ; its sign is the (right or left) polarisation. In what follows we shall choose $\kappa>0$.
Then for an arbitrary affine parameter $\sigma$ we have the null-geodesic equations,
%%%%
\begin{subequations}
\begin{align}
\frac{d^2Z}{d\sigma^2} & \; +\; C U^{-2(i\kappa + 1)} \bar{Z} = 0 \,,
\\[6pt]
\frac{d^2V}{d\sigma^2} &\;-\; 2C\left\{ \frac{\;\;d}{d\sigma}\Re\big[U^{2(i\kappa-1)}Z^2 \big]\frac{dU}{d\sigma} -\Re\big[(i\kappa-1)U^{2(i\kappa-1)}Z^2  U^{-2(i\kappa+1)}\bar{Z}^2\big]
\big(\frac{dU}{d\sigma}\big)^2 U^{-1}\right\} =  0\,.
\end{align}
\label{Zeq}
\end{subequations}
%%%
${d^2U}/{d\sigma^2} = 0$, implying that $U$ is an affine parameter itself.
Reversing the sign of the light-cone coordinates,
\beq
U \to - U, \qquad  V \to - V\,,
\label{UVsignchange}
\eeq
leaves the Lukash metric \eqref{Lmetric}, and consequently also  the equations \eqref{Zeq} invariant.
Choosing $U>0$ henceforth, we can switch to more familiar real  coordinates, in terms of which the Lukash \eqref{Lmetric}
is,
\beqa
ds^2_{L}&=&d\bX^2 + 2 dU dV \nn
\\[6pt]
&&- \left\{
\frac{C}{U^{2}}\cos\Big(2\kappa\ln(U)\Big)
\,
\big[(X^1)^2-(X^{2}]^2\big)
\,-\,
\frac{2C}{U^{2}}\sin\Big(2\kappa\ln(U)\Big)
 \,X^1X^2\right\}\,dU^2\,.
\label{BLukash}
\eeqa
The transverse equations form a complicated coupled Sturm-Liouville system,
\beq
\left(\begin{array}{cc}
(X^1)'' \\[3pt] (X^2)''
\end{array}\right)
 = -\frac{C}{U^2}
 \left(\begin{array}{rr}
 \cos (2\kappa \ln U) &- \sin (2 \kappa \ln U)
 \\[3pt]
- \sin (2 \kappa\ln U) & - \cos (2\kappa \ln U)
\end{array}\right)
 \left(\begin{array}{cc}
X^1 \\[3pt] X^2
\end{array}
\right)\,
\label{LukashBXY}
\eeq
where now $\{\,\cdot\,\}^{\prime}= d/dU$\,.

%%%%
Let us recall, for the sake of comparison,  ``ordinary'' CPP~: the transverse equations of motion of the metric \eqref{Bmetric} with profile \eqref{CPPprofile}
 are
\beq
\left(\begin{array}{cc}
\ddot{\xi} \\[2pt] \ddot{\eta}
\end{array}\right)
 =
-{C} \left(\begin{array}{rr}
\cos \omega T &+ \sin \omega T
\\[2pt]
\sin \omega T &- \cos \omega T
\end{array}\right)
\left(\begin{array}{cc}
\xi \\[2pt] \eta
\end{array}\right)\,
\label{CPPeqmot}
\eeq
where $\omega=\const$
(We changed notation, $x^1\to \xi,\, x^2 \to \eta, u \to T,\,\dot{\{\,\cdot\,\}}= d/dT $ on purpose). Having chosen $C>0$, bounded solutions arise when
\beq
\big(\frac{\omega}{2}\big)^2 - C >0\,.
\label{CPPbounded}
\eeq
Lukash and CPP are thus similar but still different.
%%%

%%%%%%%%%%%%%%%%%%%%%%%%%%%%%%%%%%%%%%%%%%%%%%%%%%%%%%%%
\subsection{Solution of the Sturm-Liouville equations}\label{SLsol}
%%%%%%%%%%%%%%%%%%%%%%%%%%%%%%%%%%%%%%%%%%%%%%%%%%%%%%%%

Sturm-Liouville equations are notoriously difficult to solve.
Analytic solutions  can be found in our case, though, by the following steps.

$\bullet$ Our clue is to switch  to ``logarithmic time'' and introduce new transverse coordinates,
\beq
\medbox{
U = e^T\,,
\quad
\bX=e^{T/2}\bxi,
\where
\bxi={\tiny \barray{c}\xi\\ \eta\earray},
}
\label{lnchange}
\eeq
in terms of which \eqref{LukashBXY} becomes,
\beq
\bigbox{
\frac{\;\; d^2}{dT^2}
\barray{c}\xi\\\eta\earray= \text{linear} + \text{CPP} =
\frac{1}{4}\barray{c}\xi\\ \eta\earray
- C
\barray{rr}
 \cos(2\kappa T) &- \sin (2\kappa T)
 \\
- \sin(2\kappa T) & - \cos (2\kappa T)
\earray
\barray{c}\xi\\[3pt] \eta\earray \,.
}
\label{lnk}
\eeq
%%%%%%%%%
 Comparison with \eqref{CPPeqmot} shows that the projected non-relativistic dynamics is that of a {repulsive linear force}
 with spring constant $\smallover{1}/{4}$, combined with a periodic ``CPP'' force. The latter may be attractive or repulsive depending on the amplitude $C>0$ and the frequency, $-2\kappa$ \cite{exactsol,POLPER,IonGW}.
%%%

$\bullet$  Our second step is to change again to new position coordinates, $\binom{\alpha}{\beta}$. The rotation with half-of-the-angle \cite{BB,POLPER} (suggested to us by Piotr Kosinski) ,
\begin{equation}
\binom{\xi}{\eta}=
R_{\kappa T} \binom{\alpha}{\beta}\,,
\quad
R_{\kappa T}  =
\left(\begin{array}{cc}
\cos \kappa T
 & + \sin \kappa T
 \\
- \sin \kappa T
& \;\cos\kappa T
\end{array}%
\right)
\label{Lukrotframe}
\end{equation}%
converts \eqref{lnk} into a  coupled Coriolis-type  system with  \emph{constant coefficients},
\vspace{-2mm}
\besub
\begin{align}
&\ddot{\alpha}
+2\kappa\, \dot{\beta}
-\Omega_-^2\, \alpha =0\,,
\\
&\ddot{\beta}
-2\kappa\, \dot{\alpha}
-\Omega_+^2 \, \beta =0\,,
\end{align}
\label{abeqmot}
\esub
where the dot $\dot{\{\,\cdot\,\}}$ means, henceforth, $d/dT$,
\beq
\Omega_-^2=\left(\kappa^{2}+\frac{1}{4}-C\right)
\aand
\Omega_+^2=\left(\kappa^{2}+\frac{1}{4}+C\right)\,.
\label{TOmegas}
\eeq
Our oscillator is anisotropic, $\Omega_+^2-\Omega_-^2=2C$ ;
the lower frequency may become negative, depending on the parameters.

\goodbreak

Remarkably, the system \eqref{abeqmot} coincides with the equations \#(14) of \BB \cite{BB}, who obtained it after a series of approximations. In the Eisenhart-Duval framework \cite{Bargmann,dgh91} it would describe a \emph{charged anisotropic linear oscillator} in the plane with frequencies $\Omega_{\pm}$\,, put into a uniform magnetic field $B=-2\kappa$.
Its behavior is  determined by the subtle competition between the  magnetic and oscillating terms, as it will be illustrated by our figures below.

$\bullet$  Our last step comes from that
the equations \eqref{abeqmot} are up to the frequency-shift $1/4$ those for a {CPP} \cite{exactsol,BB,IonGW} and can thus
be solved analytically  \cite{BB,Plyuchir,IonGW}.
We start with the  Hamiltonian and symplectic form,\vspace{-3mm}
\besub
\begin{align}
\cH &=\half\bp^{2}-\half(\Omega_-^2\alpha ^{2}+\Omega_+^2\beta^{2})\,,
\label{abHam}
\\[2pt]
\sigma &= dp^{i}\wedge d\alpha^{i} - \kappa\,\varepsilon^{ij}d\alpha^{i}\wedge d\alpha^{j}\,
\label{abSymp}
\end{align}%
\label{abHamSymp}
\esub
and introduce four phase-space coordinates
$\,w_{\pm}^1,w_{\pm}^2$ by setting%
\besub
\begin{align}
\label{polperdd1}
p^{1}&=\mu_{+} w_{+}^{2}+\mu_{-}w_{-}^{2}\,
\hskip20mm
p^{2}=-\nu_{+} w_{+}^{1}-\nu_{-} w_{-}^{1}\,,
\\
\alpha&=w_{+}^1\;+\;w_{-}^1\,,
\hskip26mm
\beta\, =w_{+}^2\;+\;w_{-}^2\,.
\label{polperdd2}
\end{align}
\label{polperdd}
\esub
Then choosing the coefficients as,
\besub
\begin{align}
%\label{mu+}
&\mu_+=\dfrac{1}{2\kappa}
\left(\sqrt{C^{2}-\kappa^{2}} -2\kappa^{2}-C\right)\,,
%\\
%\label{nu+}
&\nu_+=\;\;\dfrac{1}{2\kappa}
\left(\sqrt{C^{2}-\kappa^{2}} -2\kappa^{2}+C\right)\,,\,
\label{mu+nu+}
\\[6pt]
%\label{mu-}
&\mu_{-}=-\dfrac{1}{2\kappa}
\left(\sqrt{C^{2}-\kappa^{2}} +2\kappa^{2}+C\right)\,,
%\\
%\label{nu-}
&\quad\,\nu_{-}=-\dfrac{1}{2\kappa}
\left(\sqrt{C^{2}-\kappa^{2}} +2\kappa^{2}-C\right)\,,
\label{mu-nu-}
\end{align}
\label{munu}
\esub
decomposes the system  into two uncoupled 1D Hamiltonian systems with opposite relative signs,
\beq
\sigma =\sigma_+ -\sigma_-\,,
\qquad
H = H_+ - H_-\,,
\eeq\vskip-4mm
\noindent
where
\besub
\begin{align}
%\label{chirSymp+}
\sigma_{+} &= -\dfrac{\sqrt{C^{2}-\kappa^{2}}}{\kappa}\; dw_{+}^{1}\wedge dw_{+}^{2}\,,
\qquad
H_{+} = \;\;\dfrac{\sqrt{C^{2}-\kappa^{2}}}{2\kappa}
\left[\nu_+\, w_{+}^{1}w_{+}^{1}
+\mu_+ \, w_{+}^{2}w_{+}^{2}
\right]\,,
\label{chirHam+}
\\[8pt]
\sigma_{-} &= -\dfrac{\sqrt{C^{2}-\kappa^{2}}}{\kappa}\; dw_{-}^{1}\wedge dw_{-}^{2}\,,
\qquad
H_{-} = \dfrac{\sqrt{C^{2}-\kappa^{2}}}{2\kappa}
\left[\nu_{-}\, w_{-}^{1}w_{-}^{1}
+\mu_{-}\,\,,w_{-}^{2}w_{-}^{2}
\right]\,,
\label{chirHam-}
\end{align}
\label{chirSympHam+-}
\esub
respectively.\goodbreak

%%%%%%%%%%%%%%%%%%%%%%%%%%%%%%%%%%%%%%%%%%%%%%%
\kikezd{Strong but slow perturbation: $ C  > \kappa$}\label{C>k}\,.
%%%%%%%%%%%%%%%%%%%%%%%%%%%%%%%%%%%%%%%%%%%%%%%

For  $C > \kappa$\, $\sigma_{\pm}$ and $H_{\pm}$ in \eqref{chirSympHam+-} are real and the Hamiltonian system is regular.
The associated equations of motion are
\besub
\begin{align}
%\label{w+1eq}
\ddot{w}_{+}^{1}+\lambda_{+}^2\,
 w_{+}^{1}=0\,,
%\\
\qquad
\ddot{w}_{+}^{2}+\lambda_{+}^2 \,
 w_{+}^{2}=0\,,
%\label{w+2eq}
\label{w+12eq}
\\
%\label{w-1eq}
\ddot{w}_{-}^{1}+\lambda_{-}^2\, w_{-}^{1}=0\,,
%\\
\qquad
\ddot{w}_{-}^{2}+\lambda_{-}^2 \,
 w_{-}^{2}=0\,,
%\label{w-2eq}
\label{w-12eq}
\end{align}
\label{w+-12eq}
\esub
where the two effective frequency-squares are,
\beq
\lambda_{+}^2= \mu_{+}\,\nu_{+}=\kappa^{2}-\smallover{1}/{4} - \sqrt{C^{2}-\kappa^{2}}
\quad\aand\quad
\lambda_{-}^2= \mu_{-}\, \nu_{-}=\kappa^{2}-\smallover{1}/{4} + \sqrt{C^{2}-\kappa^{2}}\,.
\label{lambdasquares}
\eeq
%%%%%%
The equations \eqref{w+-12eq} are not independent, though: $w_{+}$ and $w_{-}$ have to satisfy
$
\dot w^1_{+} = \mu_{+} w^2_{+}$
and
$
\dot w^1_{-} = \mu_{-} w^2_{-}\,,
$
which eliminates half of the integration constants and
 we end up with two independent oscillations with frequencies
 $\lambda_{\pm}$,
\besub
\begin{align}
w_{+}^{1}&=a\cos\lambda_{+}T+b\sin \lambda_{+}T,~~~~~
w_{+}^{2}=\;\sqrt{\frac{\nu_{+}}{\mu _{+}}}\left( -a\sin \lambda _{+}T+b\cos \lambda_{+}T\right)\,,
\label{w+}
\\[4pt]
w_{-}^{1}&=c\cos\lambda_{-}T+d\sin\lambda_{-}T,~~~~~
w_{-}^{2}=-\sqrt{\frac{\nu_{-}}{\mu_{-}}}\left(c\sin\lambda_{-}T-d\cos\lambda_{-}T\right)\,,
\label{w-}
\end{align}
\label{wsol}
\esub
where $a, b, c, d$ are free real constants.
These equations show that bounded solutions arise when both lambdas are real, as illustrated below in figs.\ref{boundfig},\ref{Ccritfig} and \ref{C>kfig}.
In view of \eqref{polperdd2} the chiral decomposition can be interpreted as follows: in the rotating frame the transverse-space trajectory is decomposed into a sum,
\beq
\barray{c}\alpha(T)\\ \beta(T)\earray
=
\barray{c}w_+^1(T)\\ w_+^2(T)\earray
+
\barray{c}w_-^1(T)\\ w_-^2(T)\earray\,,
\eeq
where the first term represents a sort of ``guiding center'' and the second describes a sort of ``epicycle'' around it. Proceeding backwards, from \eqref{Lukrotframe} we infer
$
\binom{\xi}{\eta}=R_{(\kappa T)}\binom{\alpha}{\beta}
$
and then  $\bX(U)$ can be deduced from \eqref{lnchange}.
%%%%%%
Bounded $\binom{\alpha}{\beta}$ [or  $\binom{\xi}{\eta}$] motions arise
when both frequency-squares are positive, which happens when
$
\lambda_\pm^2 >0$ that requires
\beq\medbox{
\kappa <  C  < C_{crit}=\kappa^2+\smallover{1}/{4}
\with \kappa > 1/2,\,}
\label{boundcond}
\eeq
studied before by Siklos \cite{SiklosAll}  and
illustrated in fig.\ref{BBfig} below, strongly reminiscent of fig.1 in \cite{BB}. See also  \cite{PaulRMP,Kirillov}.
%%%%
Expressed in Brinkmann terms
\beq
\bX(U)=\sqrt{U}\,
R_{(\kappa \ln U)}\binom{\alpha}{\beta}
\label{BXU}
\eeq
and therefore the trajectories escape. Boundedness/unboudness will be further discussed in sec.\ref{bubSec}.

The metric admits a 3-parameter group of motions acting on spacelike hypersurfaces if and only if either
$\kappa\leq \half$ and $0 \leq  C  \leq \kappa^2+\smallover{1}/{4}\,,
$
or
$
\kappa\geq \half$ and $0 \leq  C  \leq \kappa\,
$
\cite{SiklosAll}.
The group type (with some overlaps) is~:
\begin{itemize}
\item Bianchi VII: if
\beq
\text{either}\quad
0 \leq  C  < \kappa \quad\text{or}\quad  C  = \kappa > \half
\eeq
\item
Bianchi VI: if
\beq
\text{either}\quad
\kappa <  C  \leq \kappa^2+\smallover{1}/{4}
\quad\text{or}\quad
0 <  C  = \kappa < \half
\eeq
\item
Bianchi IV: if
\beq
 C  = \kappa
\label{LLvalue}
\eeq
\end{itemize}
It follows that bound motion arise only when the wave is of Bianchi-VI  type.

Below we study the behavior for various values of the parameters.
%%%%%%%%%%%%%%%%%%
\begin{figure}[h]
\includegraphics[scale=.2]{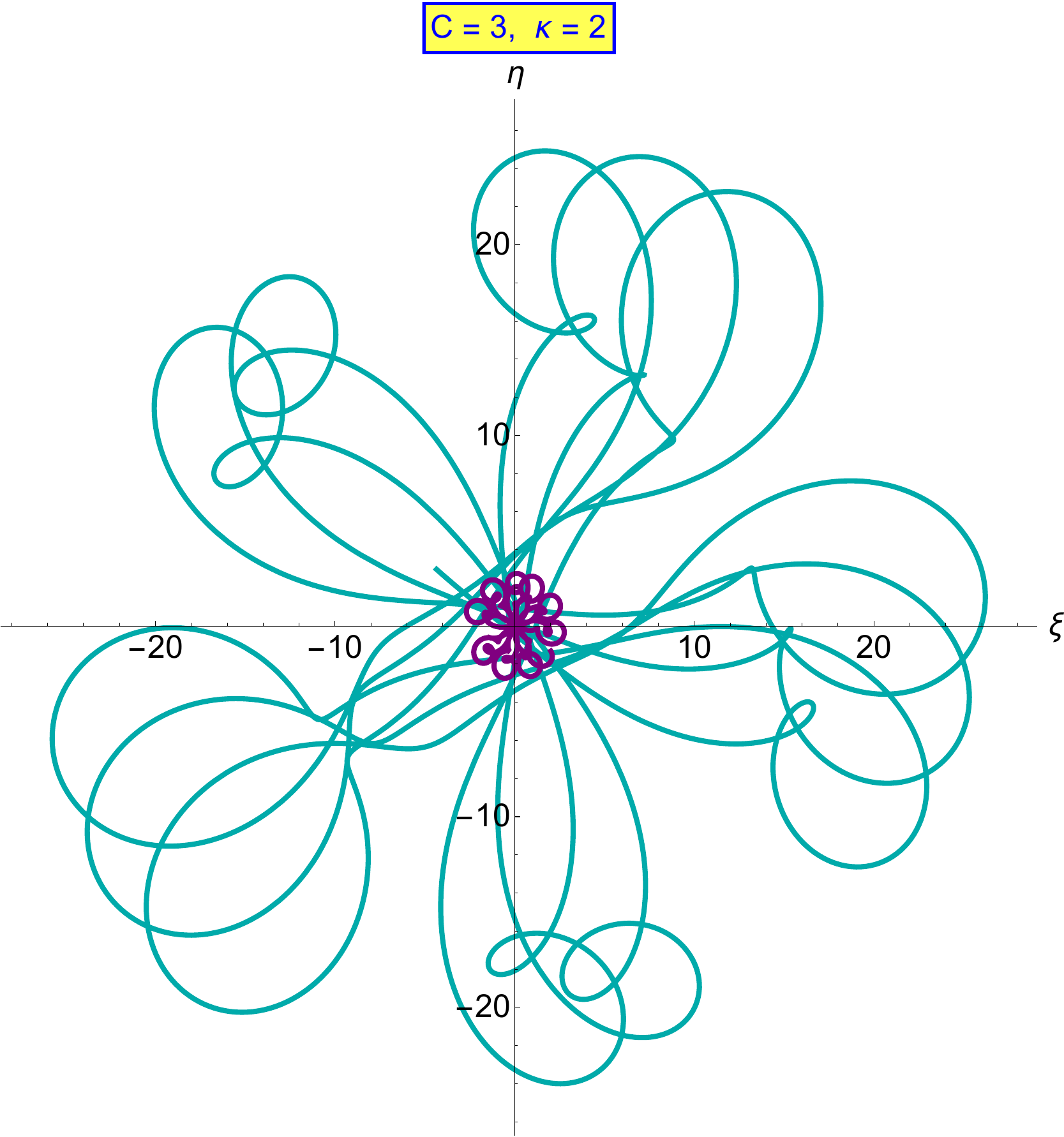}
\vskip-3mm
\caption{\textit{\small In terms of logarithmic position coordinates $\binom{\xi}{\eta}$  \eqref{lnchange} (or their rotated versions $\binom{\alpha}{\beta}$  \eqref{Lukrotframe}) one gets bounded \cyan{\bf Lukash trajectories} in the parameter range \eqref{boundcond}.
The \magenta{\bf magenta} curve near the origin is the contribution of the  ``\magenta{CPP}'' term in \eqref{lnk}, which is pushed outwards by the  repulsive force. However the \magenta{CPP} term keeps the motion bounded.
}
\label{BBfig}
}
\end{figure}
%%%%%%%%%%%%%%%%%%%
\goodbreak

%%%%%%%%%%%%%%%%%%%
\begin{figure}[h]\hskip-1mm
\includegraphics[scale=.18]{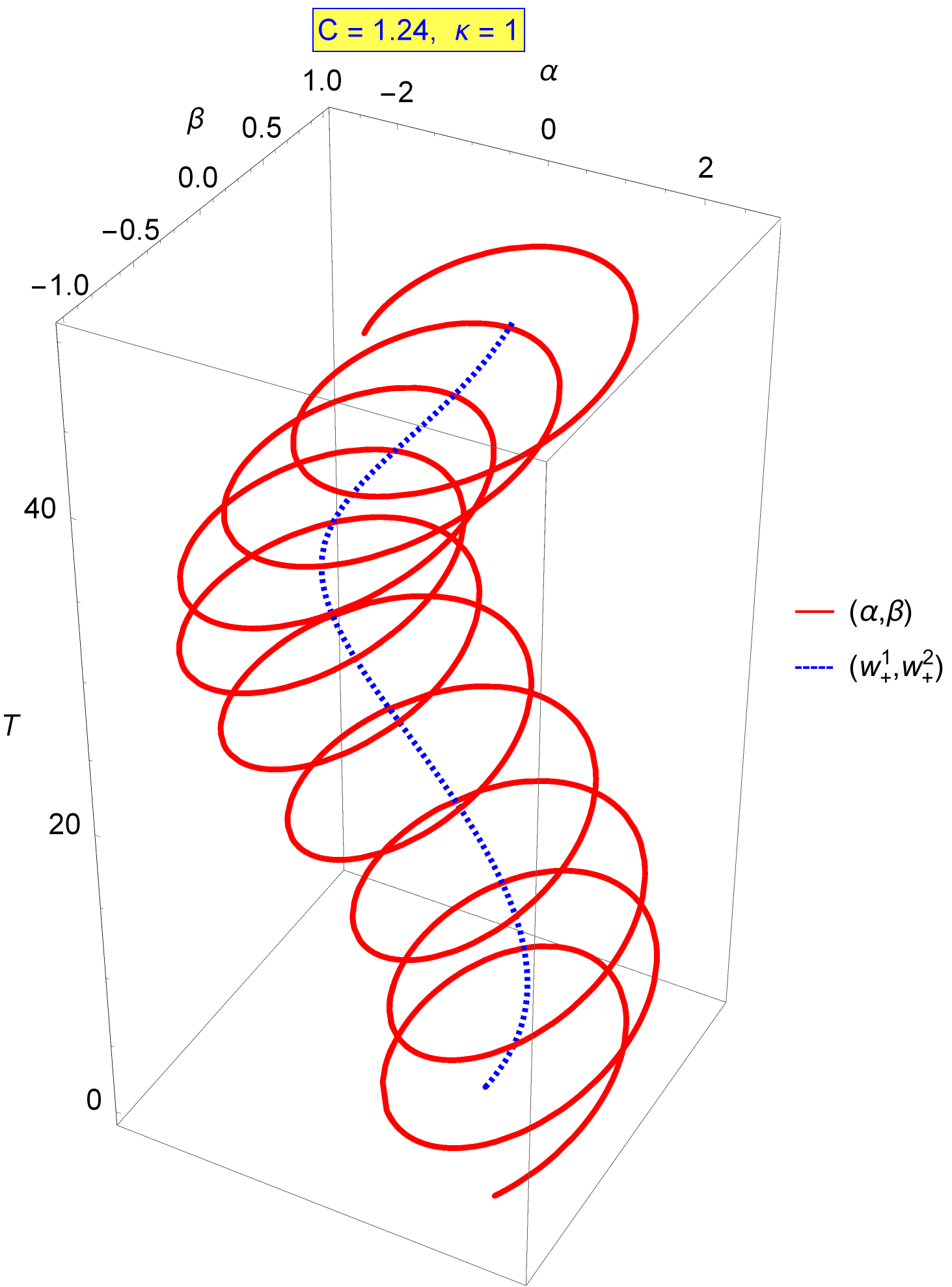}\;\;
\includegraphics[scale=.17]{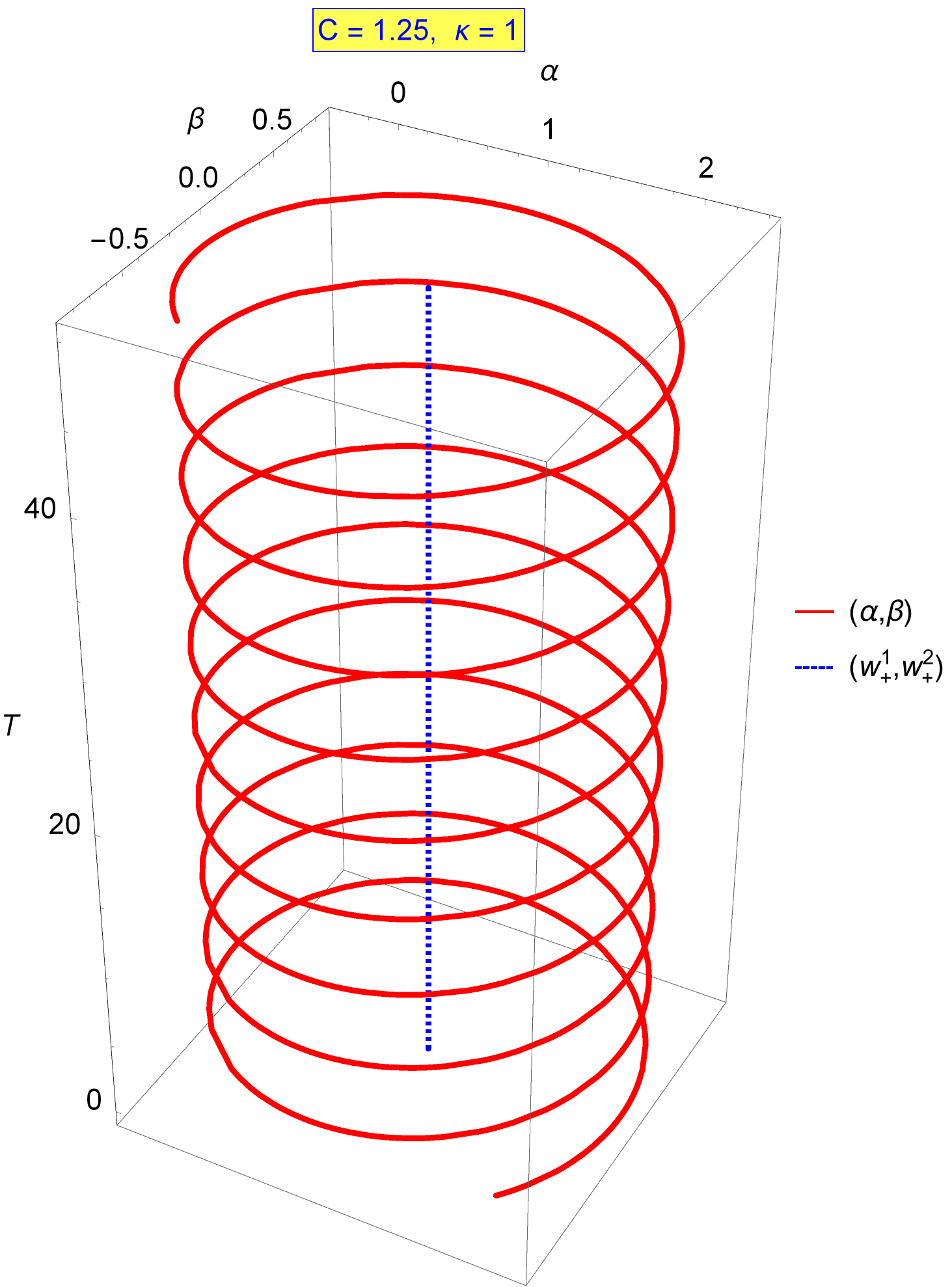}\;\;
\includegraphics[scale=.18]{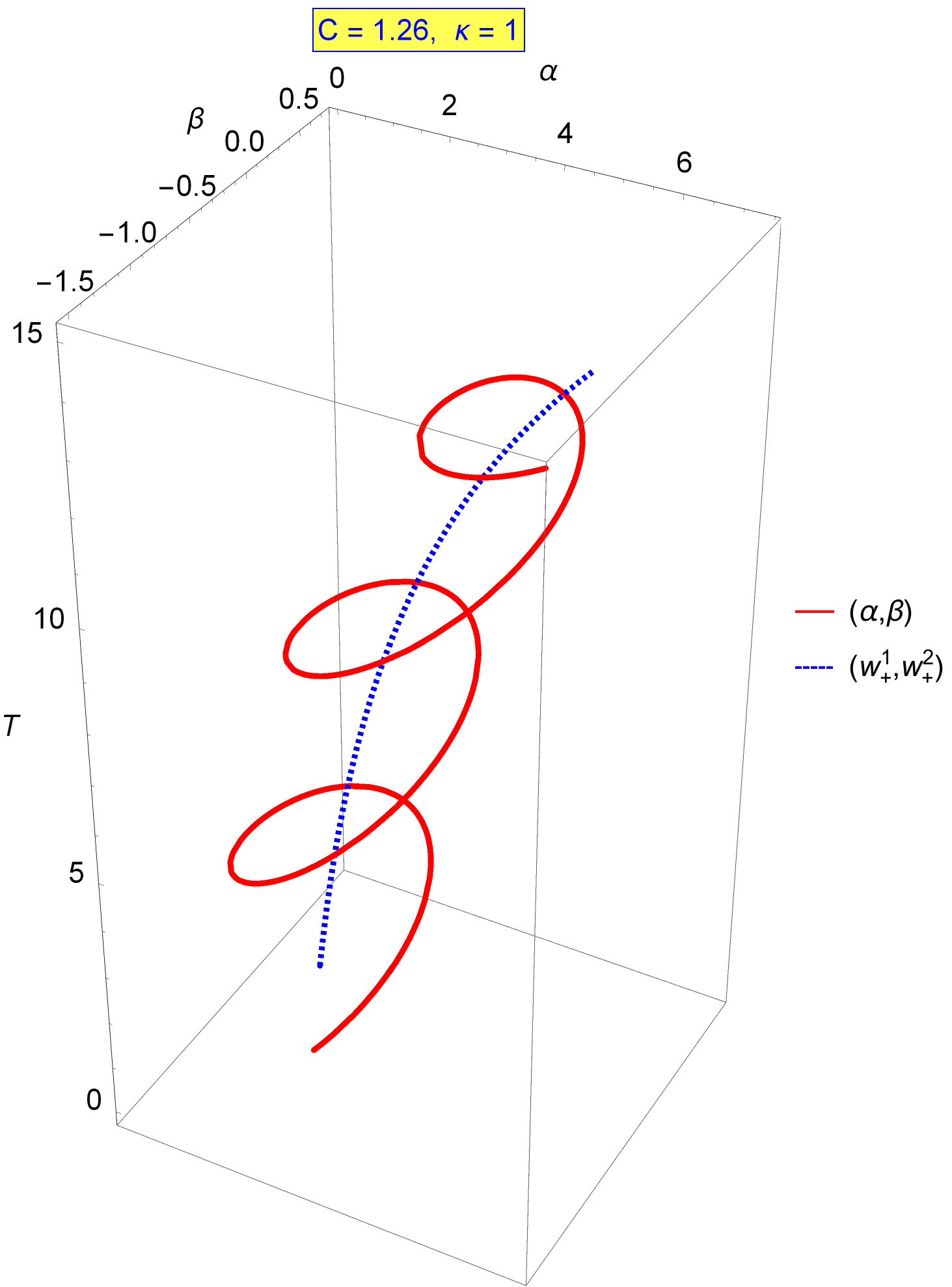}\\
%\vskip-1mm %\vspace{-6mm}
\hskip-2mm
(a) \hskip39mm (b) \hskip41mm (c)
 \vskip-3mm
\caption{\textit{\small Trajectories  shown in \red{$\binom{\alpha}{\beta}$}  coordinates unfolded in logarithmic time $T$ when close to the upper critical value $C_{crit}$. The motion of the \blue{\bf guiding center} ${\bw}_{+}$ is determined by \blue{$\lambda_+^2$} and the ``epicycle'' around it is described by $\bw_{-}$, determined by $\red{\lambda_-^2}=\lambda_+^2+2\sqrt{C^{2}-\kappa^{2}}$.
(a) when $\kappa<C<C_{crit}=\kappa^2+\smallover{1}/{4}$  (which has Bianchi type VI), the trajectory  remains bounded ;
(b) for  $C=C_{crit}$ with $\kappa > \half$  (which is Bianchi VI), the radius is constant in logarithmic time $T$ (grows linearly in $U$),  (c)  for $C>C_{crit}$ the trajectory escapes exponentially.
\label{boundfig}
}}
\end{figure}\vskip2mm
%%%%%%%%%%%%%%%%%%%%%%

\null\vskip8mm
%%%%%%%%%%%%%%%%%%%%%
\begin{figure}[h]\hskip-2mm
\includegraphics[scale=.3]{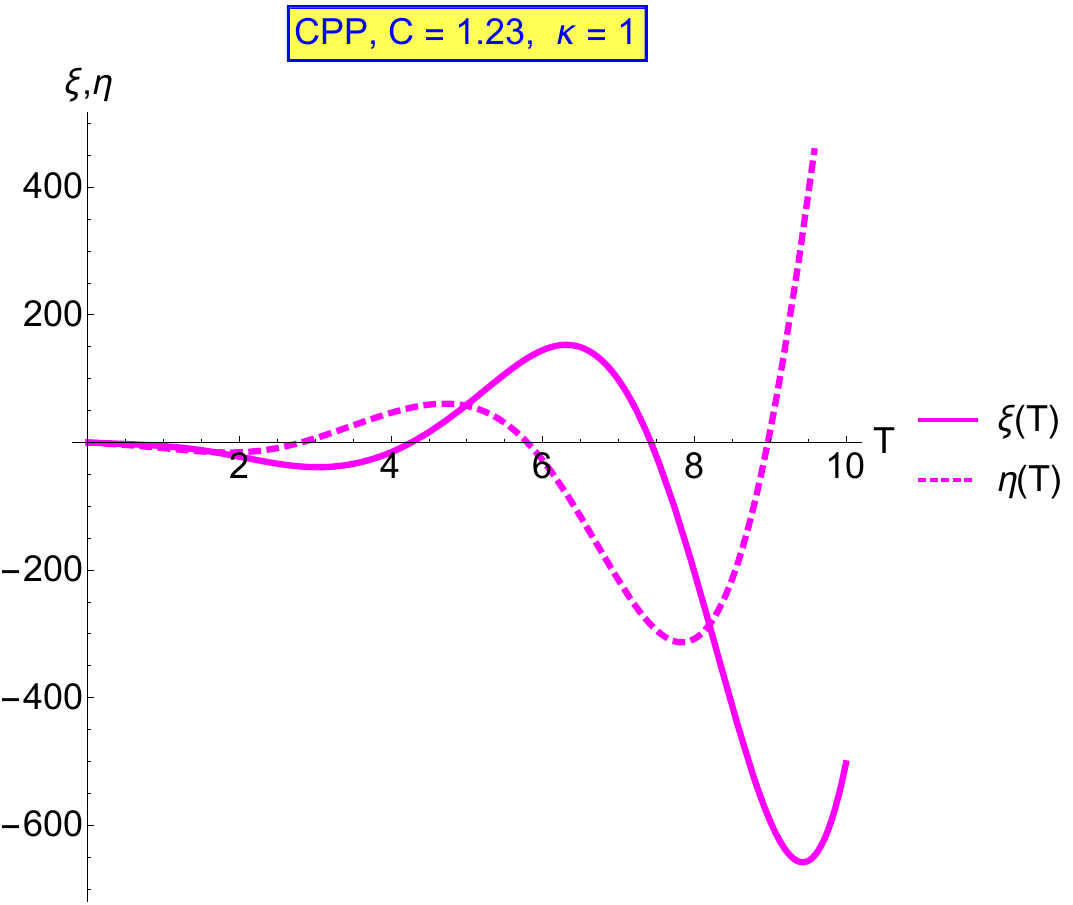}\;\;\;
\includegraphics[scale=.3]{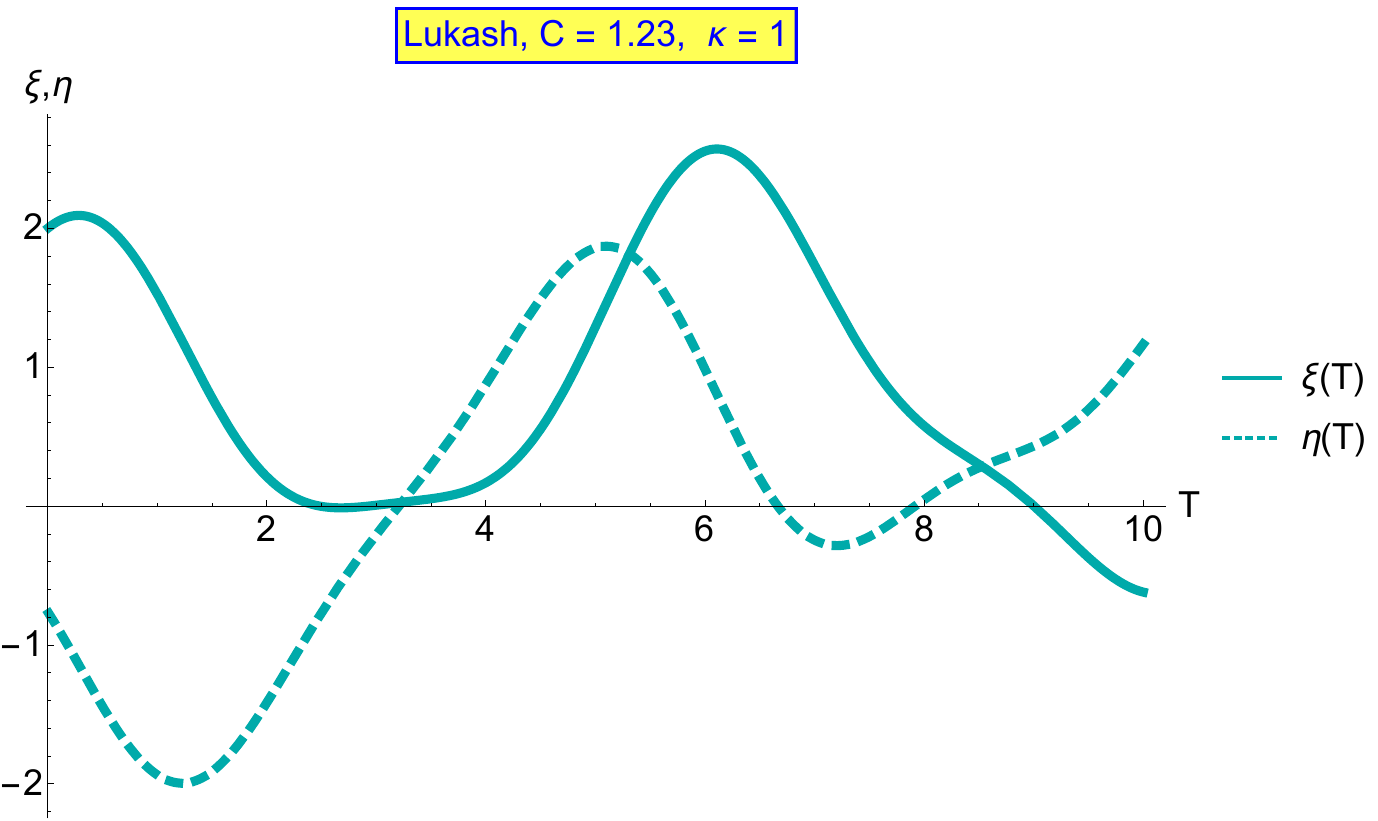}\;
\includegraphics[scale=.29]{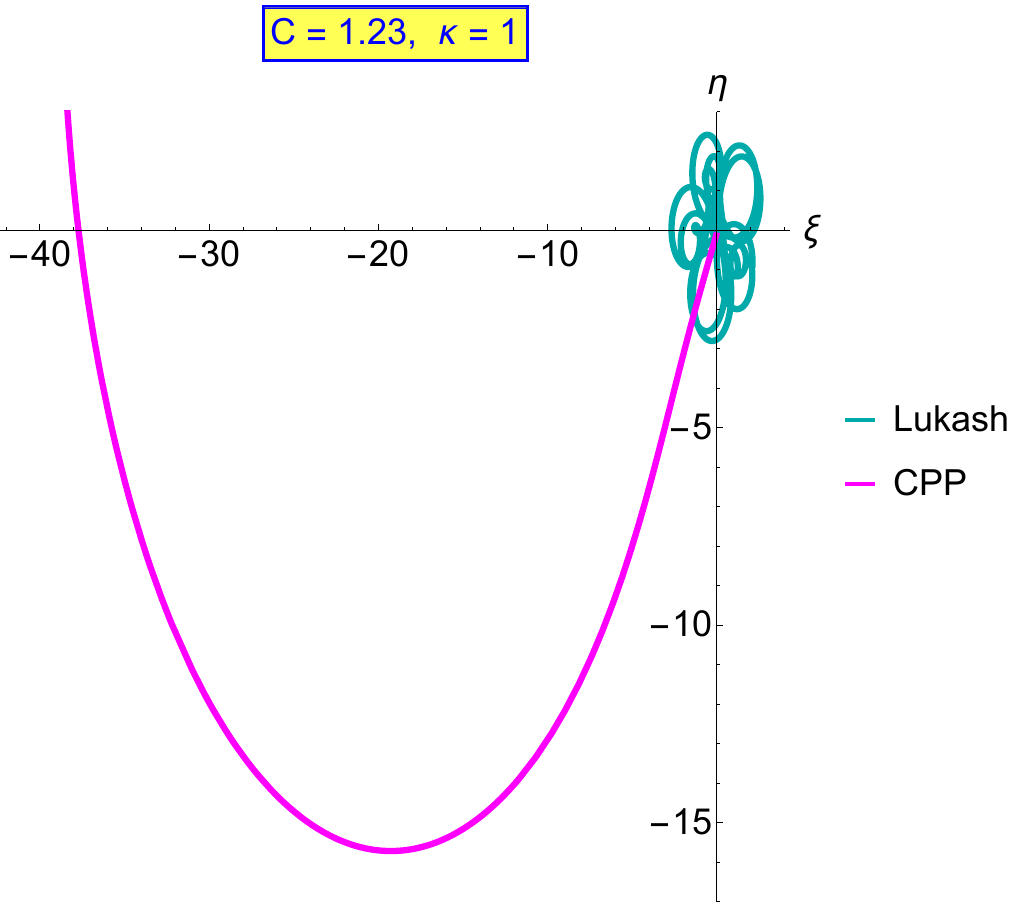}
\vskip-2mm
\hskip-6mm \magenta{(a)} \hskip55mm \cyan{(b)} \hskip57mm (c)
\\
\vskip-3mm
\caption{\textit{\small
Transverse trajectories in  $(\xi,\eta)$  coordinates just below the \underline{upper critical value} $C_{crit}$ for: \magenta{(a)} the \magenta{\bf CPP} term alone, i.e., without the linear perturbation;
 \cyan{(b)} the full \cyan{\bf Lukash} system.  \magenta{(a)}  and \cyan{(b)}  show the components. Caveat: the \magenta{CPP} and \cyan{Lukash} scales in (a) and (b) repectively are different. (c) shows the transverse trajectories.
The motion is bounded for \cyan{Lukash} but unbounded for \magenta{CPP}.
}
\label{Ccritfig}
}
\end{figure}

\vskip-2mm
\kikezd{Behaviour at the critical values}.
%%%%%%%%%%%%%%%%%%%%%%%%%%%%%%%%%%%%%%%%%%

At the \emph{upper critical  value}  $C=C_{crit}$ the effective frequencies  \eqref{lambdasquares} become
 $\lambda_+=0$ and $\lambda_-^2=2(\kappa^2-1/4)$. The motion is unbounded,
$
 w_+^1=a+b\,T,\, w_+^2=c+d\,T\,,
$
 while $w_-$ either oscillates or escapes, depending on $\kappa$ being above or below $1/2$. The behavior for $C$ close to  $C_{crit}$ is illustrated in figs. \ref{boundfig}, \ref{Ccritfig} \ref{C>kfig}.
 For $C>C_{crit}$ the motion escapes rapidly \footnote{
 The motion can remain bounded even for
$C > C_{crit}$, though. If $\lambda_+^2<0$ but $\lambda_-^2>0$ then  the trigonometric functions in  \eqref{w+} become hyperbolic,
while those in  \eqref{w-} remain oscillatory. However when $b=-a$ the  terms  $w_+^i \propto \exp(-\lambda_+T)~(i=1,2)$
fall off exponentially and the bounded (oscillatory) terms dominate.
}.
\goodbreak

\begin{figure}[h]\hskip4mm
\includegraphics[scale=.23]{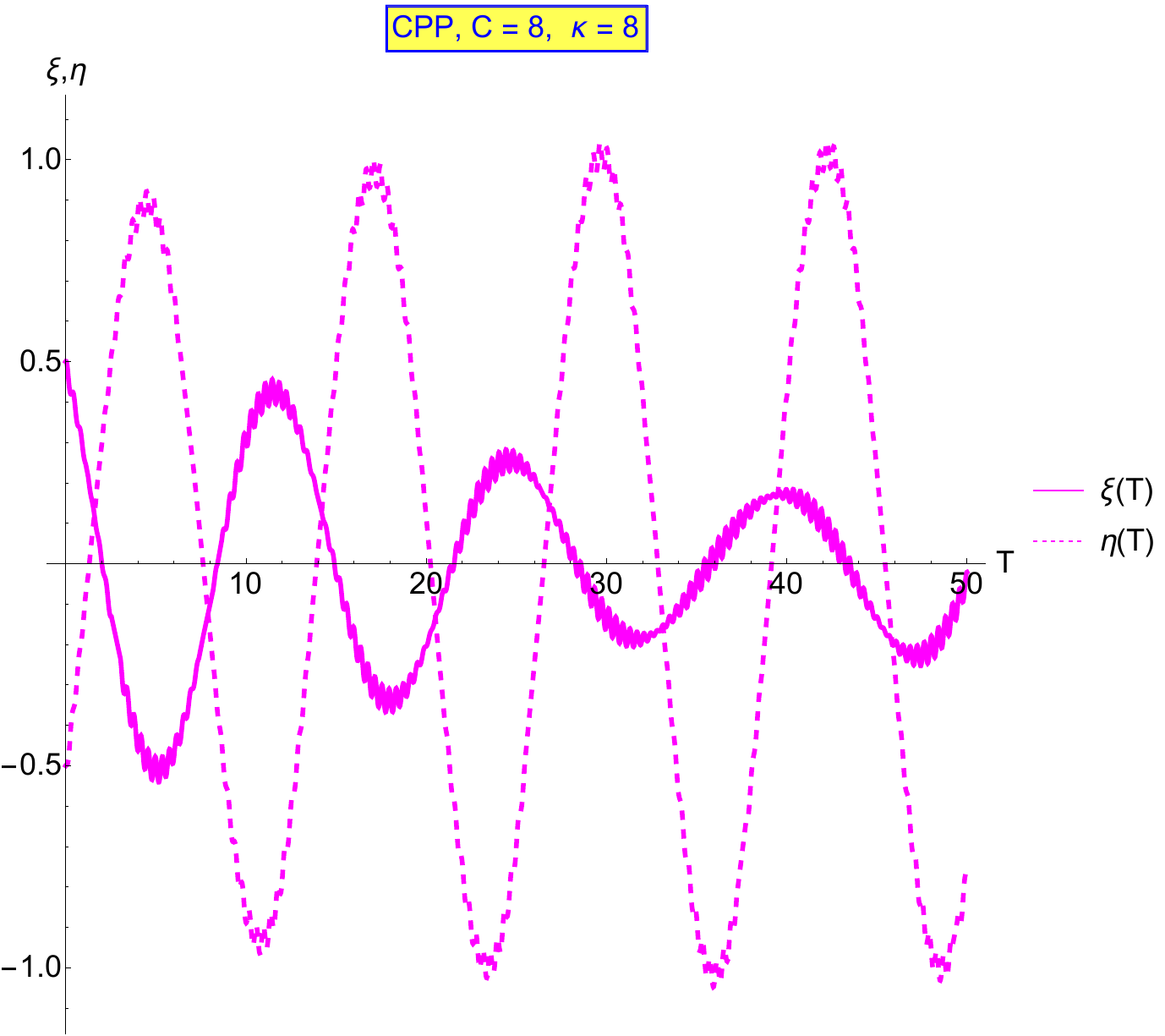}\;\;\;
\includegraphics[scale=.23]{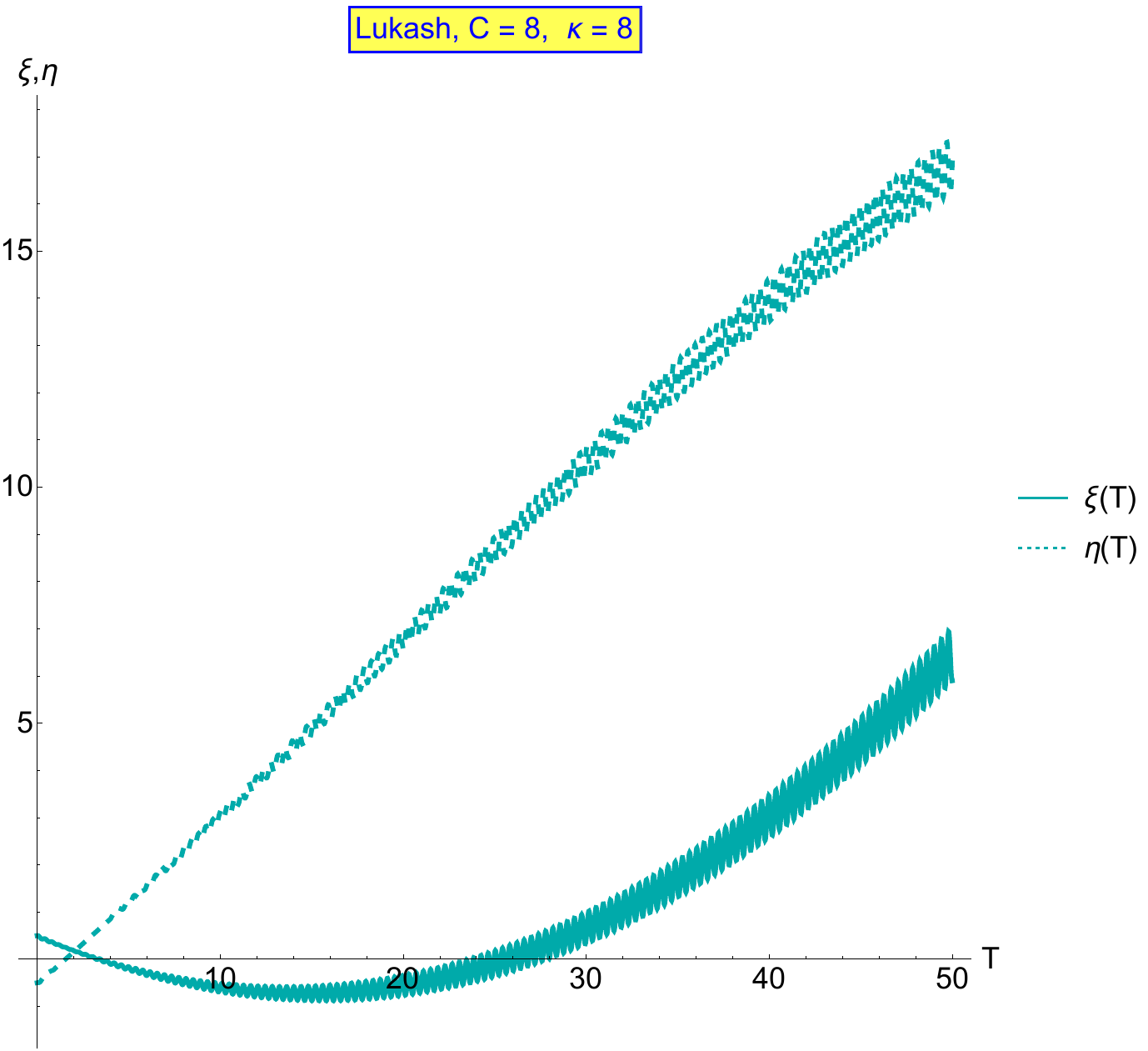}\;\;
\includegraphics[scale=.23]{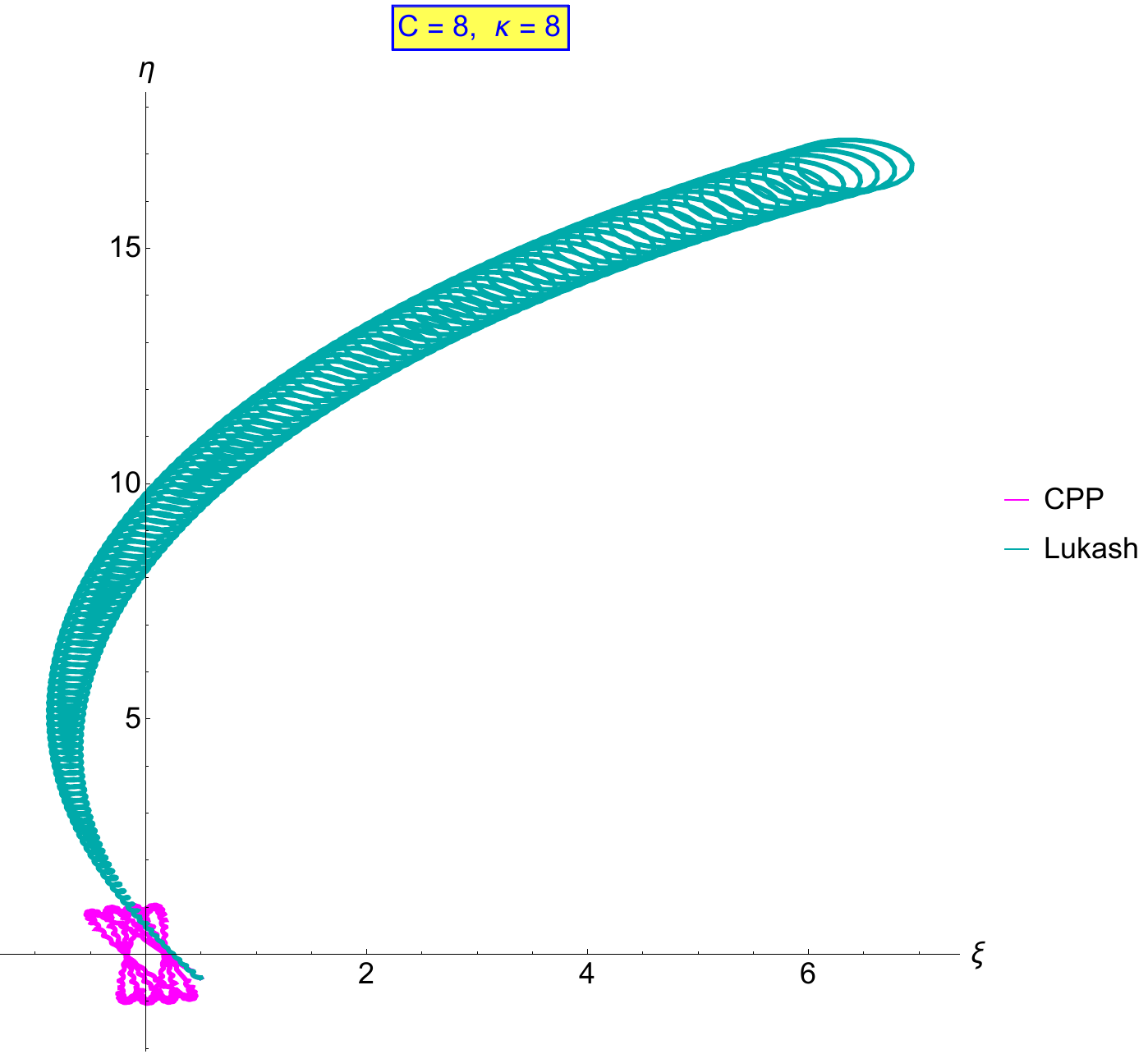}\\
\vskip-2mm
%\vspace{-1mm}
\hskip-4mm \magenta{(a)} \hskip54mm \cyan{(b)} \hskip50mm (c)
%%%%%%%%%%
\vskip-2mm
\caption{\textit{\small The components of \magenta{(a)} \magenta{\bf CPP}  and \cyan{(b)}  \cyan{\bf Lukash}  geodesics \underline{at} the \underline{lower critical value} $C = \kappa$ (which is Bianchi VII and Bianchi IV).
The colors refer to the respective cases. Caveat: the \magenta{CPP} and \cyan{Lukash} scales in (a) and (b) are different.  Fig.(c)  shows both trajectories in the transverse plane.
}
\label{C=kfig}
}
\end{figure}

%%%%%%%%%%%%%%%%%%%%%%%%%%%%%%%%%%
At the \emph{lower limiting value}
$C =\kappa$ in \eqref{LLvalue}
%%%%%%%%%%%%%%%%%%%%%%%%%%%%%%%%%%
both the symplectic forms and Hamiltonians \eqref{chirSympHam+-} vanish and we have to return to  the equations \eqref{abeqmot} with
$\Omega_{\mp}^2=(\kappa\mp\half)^{2}$.
 Then we consider three subcases, illustrated in figs.\ref{C=kfig} and \ref{C>kfig}.%
 \vskip-2mm%
\begin{itemize}
\item
For %
$
0< C=\kappa <\frac{1}{2}\,
$
(which is Bianchi VI) we get in the rotating coordinates $\binom{\alpha}{\beta}$ in \eqref{Lukrotframe},
\beqa
\alpha =&&\left(a-\kappa\big(\frac{2b+(1+2\kappa)c
}{1-2\kappa}\big)T\right) \cosh\bar{\omega}T\,+
\\[4pt]
&&\left(\frac{b+(1+2\kappa)\kappa c}{1-2\kappa}
-\frac{\kappa}{2}\big((1-2\kappa) a+2 d\big) T \right)\frac{\sinh \bar{\omega}T}{\bar{\omega}}\,,\nn
\\[12pt]
\beta =&&\left(c+\kappa \big(\frac{(1-2\kappa)a+2d}{1+2\kappa}\big) T\right) \cosh \bar{\omega}T\,+
\\[4pt]
&&\left(\frac{-(1-2\kappa)\kappa a+d}{1+2\kappa}+\kappa\big(b+(1+2\kappa)2c\big) T\right) \frac{\sinh \bar{\omega}T}{\bar{\omega}}\,,
\eeqa
where\vspace{-2mm}%
\begin{equation}
\bar{\omega}=\sqrt{\smallover{1}/{4}-\kappa^{2}}\,.
\label{k<omega}
\end{equation}
The trajectory is  exponentially escaping: the rotation is too week to keep the motion bounded.
\item
For
$
 C =\kappa >\frac{1}{2}\,
$
(which is Bianchi VII) we get the same formulae \emph{up to replacing  hyperbolic functions by trigonometric ones}, \vspace{-2mm}%
\beq
\cosh\bar{\omega}\to \cos\tilde{\omega}\,,
\quad
\sinh\bar{\omega} \to \sin\tilde{\omega}\,
\where
\tilde{\omega}=\sqrt{\kappa^{2}-\smallover{1}/{4}}\,,
\label{k>omega}
\eeq
see fig.\ref{C=kfig}.
Exponential escape is eliminated, however terms which are linear in $T =\ln U$ may remain.

\item
In the Bianchi IV case
$
 C =\kappa =\frac{1}{2},
$
the trajectories are polynomial functions of $T$,
\beq
\alpha
= a+bT -\frac{d}{2}T^{2}-\smallover{(b+c)}/{6}T^{3},
\;
\beta
=c+dT+\smallover{(b+c)}/{2}T^{2}
\,,
\eeq
which remain bounded only in the trivial case $b=c=d=0$.

\end{itemize}
%%%%%%%%%%%
The  value \eqref{LLvalue} separates our present domain of investigations  \eqref{boundcond} from  the Bianchi VII${}_{h}$ range
\beq
0 <  C <\kappa\,
 \label{BVII}
\eeq
we studied  in \cite{Lukash-I} using a rather \emph{different} (``Siklos'' \cite{SiklosAll}) technique. The results are consistent, though.
 In the range \eqref{BVII} the chiral decomposition \eqref{chirSympHam+-} would yield imaginary symplectic forms and Hamiltonians. %%%

%%%%%%%%%%%%%%%%%%
%%%%%%%%%%%%%%%%%%
%%%%%%%%%%%%%%%%%%
\begin{figure}[h]\hskip-2mm
\includegraphics[scale=.23]{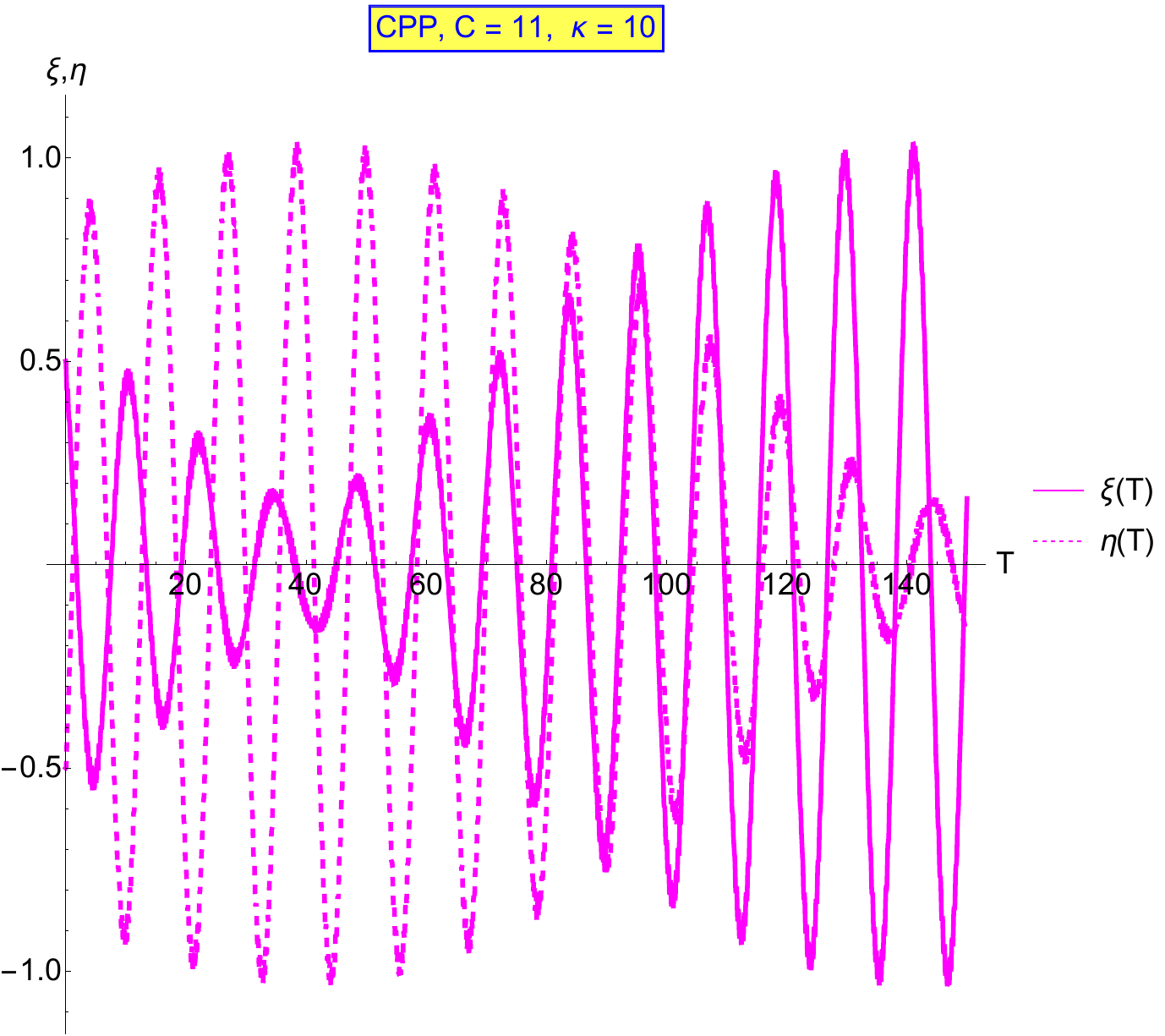}\;\;\;
\includegraphics[scale=.23]{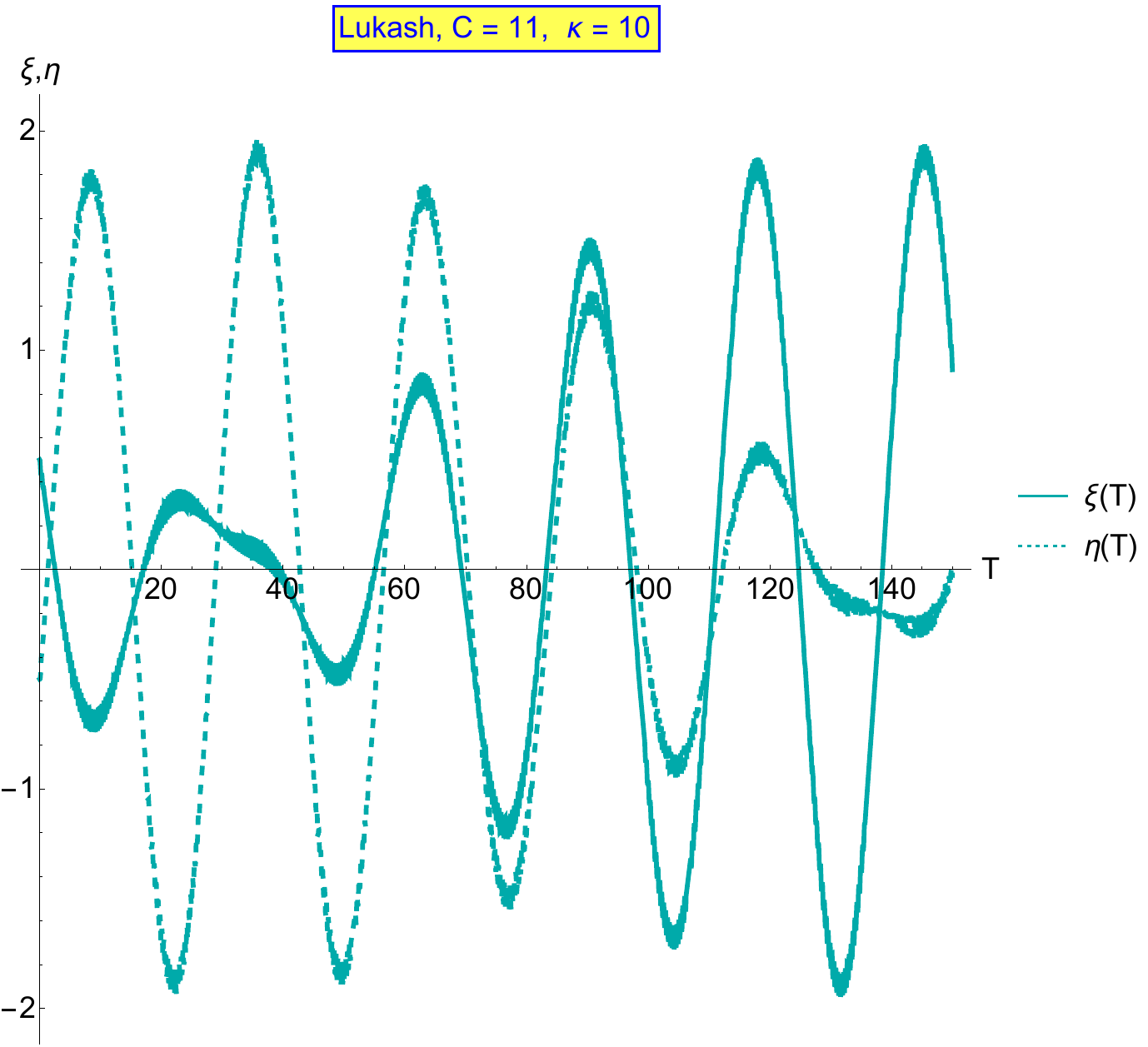}\;\;
\includegraphics[scale=.245]{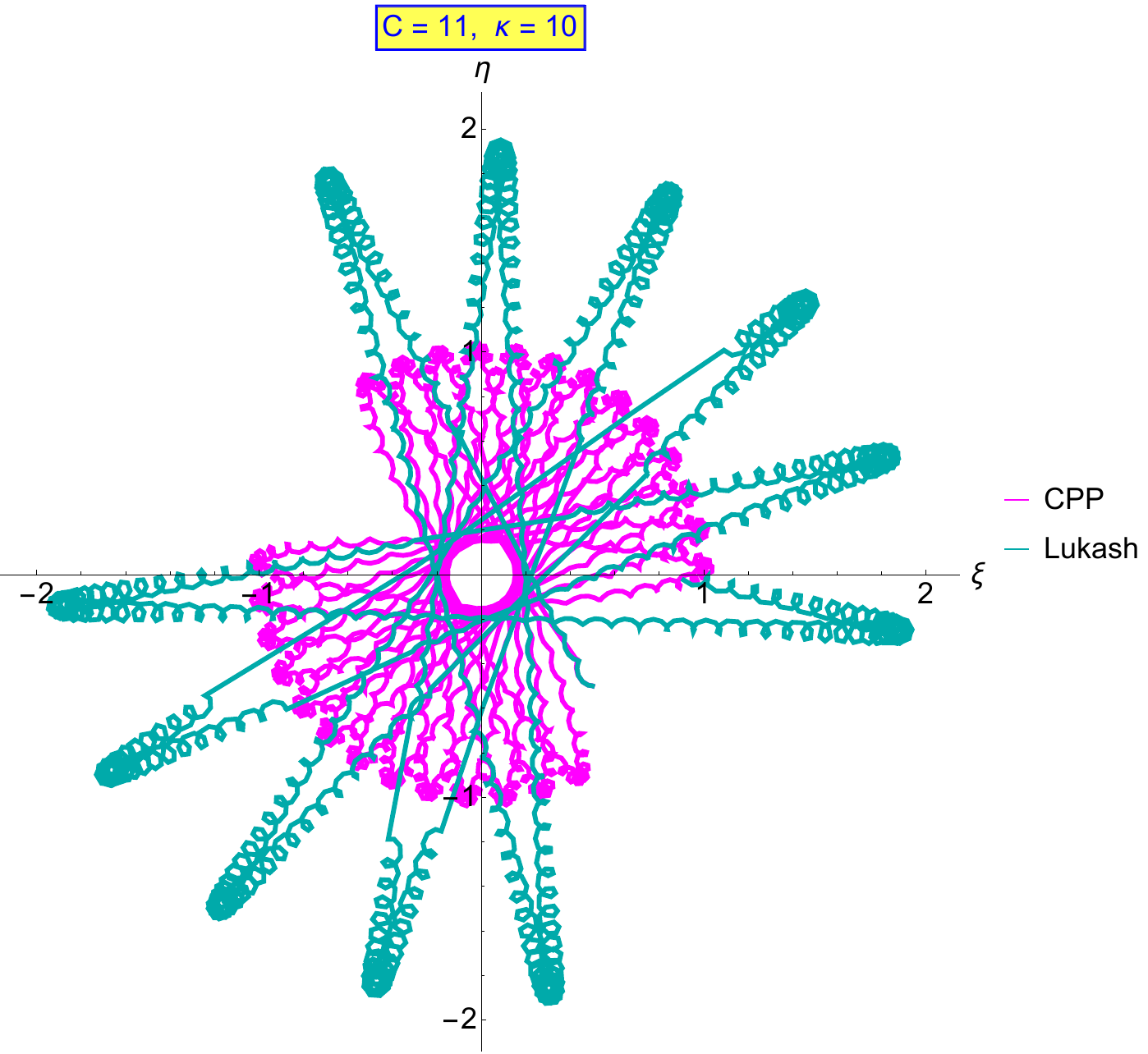}\\
\vskip-3mm
\hskip-10mm \magenta{(a)} \hskip50mm \cyan{(b)} \hskip50mm (c)
\\
\vskip-4mm
\caption{\textit{\small  \magenta{(a)} \magenta{\bf CPP} and  \cyan{(b)}  \cyan{\bf  Lukash} trajectories \underline{just above} the \underline{lower critical value} $C = \kappa$. The colors and labels refer to the respective cases.
Caveat: the \magenta{CPP} and \cyan{Lukash} scales are different.
(c) combines the two curves.
}
\label{C>kfig}
}
\end{figure}
%%%%%%%%%%%%%%%%%%

%%%%%%%%%%%%%%%%%%%%%%%%%%%%%%%%%%%%%%
\section{Lift to 4D}\label{LiftSec}
%%%%%%%%%%%%%%%%%%%%%%%%%%%%%%%%%%%%%%

Further insight is provided by lifting the coordinate transformation to 4D Lukash spacetime.
Completing \eqref{lnchange} with $V \to V$ allows us to
write the Lukash metric as
\beq
ds_L^{2}= e^{T}d\Sigma^2\,,
\label{LconfCPP}
\eeq
%%%%%%
\vskip-3mm\noindent
which is conformal to
\beq
d\Sigma^2=d\xi^{2}+d\eta^{2}+2dTdV+({\xi} d\xi+{\eta} d\eta)dT+\big(\frac{1}{4}\left(\xi^{2}+\eta^{2}\right) -C\left(\xi^{2}-\eta ^{2}\right) \cos 2\kappa T+2C\xi \eta
\sin 2\kappa T\big) dT^{2}\,.
\label{Sigmametric}
\eeq
From the Barmann point of view, the geodesics of this metric project to motion in a \emph{uniform magnetic field} combined with an \emph{anisotropic oscillator}.
The term $({\xi} d\xi+{\eta} d\eta)dT$ can actually be absorbed by
a redefinition of the vertical coordinate,
\beq
V = \nu - \smallover{1}/{4}\big(\xi^{2}+\eta^{2}\big)\,,
\label{Vnushift}
\eeq
allowing us to present \eqref{Sigmametric} as,
\beq
d\Sigma^2=d\bxi^{2}+2dTd\nu+\big(\frac{1}{4}\bxi^{2} -C\left(\xi^{2}-\eta ^{2}\right) \cos 2\kappa T+2C\xi \eta
\sin 2\kappa T\big) dT^{2}\,,
\label{perCPP}
\eeq
which differs from the CPP metric only in an additional
$\frac{1}{4}\bxi^{2}$ perturbation of the potential (where we wrote obviously $\bxi=(\xi,\eta)$).
In conclusion,
\beq
(\bX,U,V) \to (\bxi, T,\nu)\,,
\label{XUVxiTnu}
\eeq
is a conformal mapping from Brinkmann to coordinates
with logarithmic time, $T=\ln U$. We shall call it ``perturbed CPP metric'' in what follows. The latter is not a vacuum solution due to the additional $\bxi^2$ in the potential.

%%%%%%
 Conformally related metrics have identical null geodesics (and intertwined respective affine parameters) \cite{Wald}; the identity of the null geodesics of   $ds_L^2$ and $d\Sigma^2$, respectively,  can be checked directly.

Henceforce we work with  $d\Sigma^2$  and we choose $T$ as affine parameter.

We record for later use that applying  again the rotational trick  \eqref{Lukrotframe} and putting
$
A_\alpha=\kappa\beta,\,
A_\beta=-\kappa\alpha\,
$ allows us to rewrite $d\Sigma^2$ as
\beq
d\Sigma^2 = d\alpha^{2}+d\beta^{2}
+2\Big(d\nu+A_{\alpha} d\alpha+A_{\beta} d\beta\Big)dT
+\Big(\Omega_-^2 \alpha^{2}+\Omega_+^2 \beta^{2}\Big)\,dT^{2}\,.
\label{Lukabg}
\eeq
$\half(\alpha d\alpha+\beta d\beta)=d\left(\smallover{1}/{4}\big(\xi^{2}+\eta^{2}\big)\right)$  is a vector potential for a constant magnetic field $B=-2\kappa$  which can be removed by redefining the vertical coordinate,
 cf. \eqref{Vnushift}.

%%%%%%%%%%%%%%%%%%%%%%%%%%%%%%%%%%%%%%%%%%%%%%%%%%%%%%%%%
\section{Isometries and conformal transformations}\label{IsoSect}
%%%%%%%%%%%%%%%%%%%%%%%%%%%%%%%%%%%%%%%%%%%%%%%%%%%%%%%%

Now we show that the Lukash metric carries a 6-parameter group of isometries which extends to a 7 parameter conformal group.
We follow first our road map set out in sec.\ref{SLsol}.
\benu
\item
The Lukash metric admits the generic 5-parameter broken Carroll
symmetry  of gravitational plane waves \cite{exactsol,LL,Carroll4GW,Torre,SLC,Conf4GW} spanned by the covariantly constant vector $\p_V$
plus  of $4$ ``U-dependent translations'' \cite{GiPo},
\beq
X^1\to X^1 + \beta^1(U),\qquad X^2\to X^2 + \beta^2(U)\,
\label{Udeptr}
\eeq
that carry solutions into solutions. Then by the \emph{linearity} of \eqref{ABXeq}  the $\beta^i$  \emph{have to satisfy the same equations as the transverse coordinates do}. Thus the \emph{time-dependent symmetries of \eqref{ABXeq} are the projections of geodesics} \cite{Carroll4GW}. They lift to 4D to isometries by \eqref{Vlift} and are analogous to what one obtains for an isotropic oscillator by pulling back the translations and boosts of a free particle by Niederer's map  \cite{Niederer73}. They  span a Newton-Hooke group structure \cite{GiPo,Zhang:2011bm}.

\item
The remarkable property of the Lukash system is its additional 6th isometry \cite{exactsol,Lukash-I}, recovered as follows~: in terms of the new coordinates $T = \ln U$, $\binom{\alpha}{\beta}$ and $\nu$ in \eqref{Lukrotframe}--\eqref{Vnushift}
none of the coefficients in the rotated metric \eqref{Lukabg} [or of the equations of motion \eqref{abeqmot}] depends on $T$; therefore $T$-translations,
\beq
T \to  T +\tau\,,\qquad \binom{\alpha}{\beta} \to  \binom{\alpha}{\beta}, \qquad \nu\to \nu\,
\label{Ttransl}
\eeq
are manifest isometries for any real constant $\tau$.
Working backwards, the  perturbed CPP equations \eqref{lnk} are invariant w.r.t.   the ``screw'' obtained by combining a $T$-translation with a transverse rotation with angle
$(-\kappa\tau)$ \cite{Carroll4GW,POLPER,IonGW},
\beq
T \to T+\tau,\qquad  \binom{\xi}{\eta} \to R_{(-\kappa \tau)}\,\binom{\xi}{\eta}\,,
\qquad
\nu\to \nu\,.
\label{Tscrew}
\eeq
In Brinkmann terms this isometry is implemented as
\beq
U\to e^{\tau}\, U, \qquad \bX \to  R_{(-\kappa \tau)}\bX, \qquad
V \to e^{-\tau}\,V\,
\label{muUXV}
\eeq
and is generated by combining an U-V boost with a transverse rotation \footnote{$Y_{\kappa}$ is actually a symmetry  for any $\kappa$ and $C$, and is consistent with eqn. (3.19) in  \cite{Lukash-I} valid in the Bianchi VII case.},
\beq
Y_{\kappa} =  (U\partial_U-V\partial_V)
- \kappa (X^1\partial_2 - X^2 \partial_1)\,
\label{LUVboost}
\eeq
we propose to call ``expanding screw''.
For $\kappa=0$ the Lukash profile reduces to $U^{-2}$ and the results in \cite{Conf4GW,Conf4pp} are recovered.

\item
In addition to  isometries, we also have the homotheties,
\beq
U \to U, \qquad \bX \to \lambda\,\bX,\qquad
V \to \lambda^{2} V\qquad
(\lambda =\const > 0),
\label{UHomo}
\eeq
which are conformal transformations with conformal factor
$\lambda^2$  \cite{Hsu,Conf4GW}. \eqref{UHomo} is generated by,
\beq
\label{Yhom}
Y_h = X^i\frac{\partial}{\partial X^i} + 2V \frac{\partial}{\partial V}\,.
\eeq
\eenu

Now we relate the 6th isometry \eqref{LUVboost} for Lukash   to the ``screw'', known before for CPP. The latter  combines a $T$-translation with a transverse rotation,
\beq
\label{screw1}
Y_{s} =
\partial_T + \frac{\omega}{2} \epsilon_{ij} \xi^i \partial_j \,.
\eeq

We first recall how it goes for CPP.
Using again  the notations $\bx \to \bxi=(\xi,\eta)$, $u \to T,\, v \to \nu$, the CPP metric
\eqref{Bmetric}-\eqref{Bprofile}-\eqref{CPPprofile} is written, in complex notation $\zeta = {(\xi + i\eta)}/{\sqrt{2}}$\,,
\beq
ds^2_{CPP} =
2 d\zeta d\bar{\zeta} + 2d\nu dT + A_0 \Re[e^{-i\omega T}\zeta^2] dT^2.
\label{CPP1}
\eeq
Then for any $\tau=\const$
\beq
\label{screw2}
T \to T + \tau, \quad \zeta \to e^{i \tau\omega/2} \zeta\,, \quad  \nu \to \nu
\eeq
leaves the metric \eqref{CPP1} invariant
\cite{POLPER,exactsol}. Infinitesimally, this is generated by
\beq
\label{Cscrew}
Y_{s} =
\partial_T + i\frac{\omega}{2}\big(\zeta\,\p_{\zeta}-\bar{\zeta}\, \p_{\bar{\zeta}}\big) \,,
\eeq
cf. \eqref{screw1}.
Thus CPP has a 6-parameter isometry group \cite{exactsol,BoPiRo,Carroll4GW} \footnote{The half-angle rotation \eqref{Lukrotframe} reduces \eqref{screw2} to a mere $T$-translation --- it ``unscrews the screw''.},
extended to 7-parameter conformal symmetry by the homothety
\beq
Y_h = \xi^i  \frac{\partial }{\partial \xi^i} + 2\nu \partial_{\nu}\,.
\label{Yhomxi}
\eeq

Turning to  Lukash waves, we assume $U>0$ and start with the real form  \eqref{BLukash} which is conformal to the ``perturbed CPP metric'' $d\Sigma^2$ in \eqref{Sigmametric} [or in \eqref{perCPP}],
 $ds_L^2=e^Td\Sigma^2$ cf. \eqref{LconfCPP}.
 $d\Sigma^2$ is readily seen to admit also the ``screw'' isometry
\beq
\label{Yscrewmod}
Y_s = \partial_T - \kappa\epsilon_{ij} \xi^i \partial_j,
\eeq
cf. \eqref{screw1} with $\omega \to -2\kappa$, as well as the homothety
 \eqref{Yhomxi}. These symmetries have to be pulled back to Lukash,
which involves the conformal factor $e^T$ see \eqref{LconfCPP}. However  we find
\begin{eqnarray}
L_{Y_h} \big(e^T g_{\mu\nu} (\xi, T, V)\big) &=& 2 e^T g_{\mu\nu} (\xi, T, V),
\\
L_{Y_s} \big(e^T g_{\mu\nu} (\xi, T, V)\big) &=& e^T   g_{\mu\nu} (\xi, T, V)
\end{eqnarray}
which are now both conformal.
But combining them as,
\begin{eqnarray}
Y_{\kappa} =Y_s  - \frac{1}{2} Y_h  =
 \partial_T - V\partial_V - \left(\frac{1}{2} \xi^j + \kappa \epsilon_{ij} \xi^i \right) \frac{\partial }{\partial \xi^j} ,
\label{Ykappamod}
\end{eqnarray}
the conformal factors compensate, providing us finally with
an \emph{isometry} $Y_\kappa$,
\beq
L_{Y_\kappa} \big(e^T g_{\mu\nu} (\xi, T, V) \big)= 0 .
\eeq
Then returning to our original Brinkmann coordinates by \eqref{lnchange},
the Lukash isometry \eqref{LUVboost} and the homothety \eqref{Yhom} are recovered, %
\begin{eqnarray}
Y_{\kappa} &=& U \partial_U - V\partial_V - \kappa \epsilon_{ij} X^i \partial_j,  %
\\[4pt]
Y_h &=&  X^i\frac{\partial}{\partial X^i} + 2V \frac{\partial}{\partial V}\,.
\end{eqnarray}

We note for completeness that the ``expanding screw'' \eqref{muUXV}--\eqref{LUVboost} can also be obtained along the same lines as for the CPP screw \eqref{screw2}~:
the complex forms \eqref{Lmetric}--\eqref{Zeq} are  manifestly invariant under the ``expanding screw'' transformation
\beq
U \to e^\tau U, \qquad  V \to e^{-\tau} V, \qquad
Z \to e^{-i\kappa\tau}Z\,.
\label{CLscrew}
\eeq

%%%%%%%%%%%%%%%%%%%%%%%%%%%%%%%%%%%%%%%%%%%%%%%%%%%%%%%%%%%%%%%%%
\section{Coordinate dependence of bounded/unbounded property}\label{bubSec}
%%%%%%%%%%%%%%%%%%%%%%%%%%%%%%%%%%%%%%%%%%%%%%%%%%%%%%%%%%%%%%%%%

Now, as we realised while answering a question of our referee, we argue that the \emph{very notion of ``particle trapping''}
(which appears in the title of ref. \cite{BB}, and also in the previous version of our paper) may be \emph{coordinate dependent}~: the motion can appear bounded in one coordinate system and unbounded in another one, as we illustrate it on various examples.

\benu

\item
Let us first consider the Niederer correspondence between a harmonic oscillator of frequency $\omega=\const$ and a free particle \cite{Niederer73,dgh91,Andr18,ZZH,Silagadze}. The  mapping
\beq
T = \frac{\tan\omega t}{\omega},
\qquad
X = \frac{x}{\cos\omega t}
\label{Niederermap}
\eeq
 carries the half oscillator-period $-\frac{\pi}{2\omega}  < t < \frac{\pi}{2\omega} $ into a full free motion with $-\infty < T < \infty$, whereas the bounded oscillations become unbounded.

\begin{figure}
[ht]\hskip-6mm
\includegraphics[scale=.36]{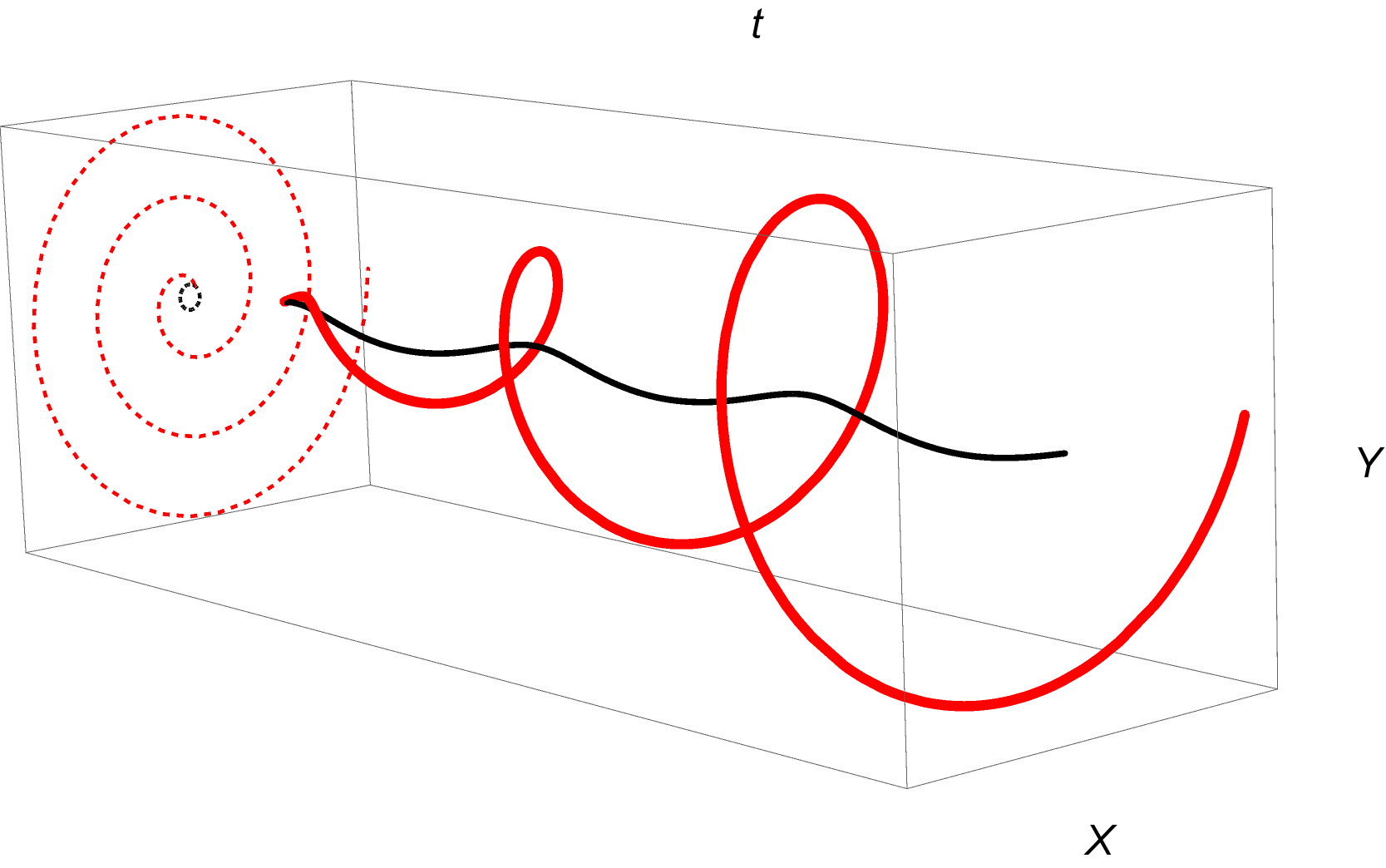}
\vskip-5mm
\caption{\textit{\small Planetary motions for gravitational constant ${\bf G_0=\const}$ and for $\red{\bf G(t)=G_0{t_0}/{t}}$, respectively, are  described by conformally related Bargman manifolds} \cite{dgh91}. \textit{\small An orbit  (in black) which is bounded for $G_0$ can become unbounded for $\red{G(t)}$, as illustrated for a circular Newtonian trajectory.
}
\label{Dtraj}
}
\end{figure}

\item
A similar behavior is observed also for planetary motion with a time-dependent gravitational constant as suggested by Dirac \cite{DiracGt}, e.g.,
\beq
G(t)= G_0\frac{t_0}{t}\,.
\label{DGt}
\eeq
For $t$ close to $t_0$, $t \approx t_0(1+\epsilon)$.
The Bargmann manifolds with $G=G_0=\const$ and $G = G(t)$, respectively, are conformally related with conformal factor $\Omega^2(t)=({t_0}/{t})^2$ \cite{dgh91}. A circular transverse trajectory $\bzeta(T)=e^{iT}$ for $G=G_0$ becomes
\beq
t \to T = -{t_0^2}/{t},\qquad
Z(t) = \frac{t}{t_0}\,e^{-i{t_0^2}/{t}}
\approx
(1+\epsilon) e^{-it_0(1-\epsilon)}.
\label{Dspiral}
\eeq
Thus for $G(t)$ in \eqref{DGt} the orbit of planetary spirals outwards, as shown in FIG.\ref{Dtraj}~:
 $G(t)$ decreases with increasing $t$. The gravitational pull weakens, the particle escapes and its rotation slows down.

 Let us note that the relativistic coordinate transformation \eqref{Dspiral} lifted to Bargmann space transforms the two different underlying non-relativistic systems with $G(t)$  and $G_0$ into each other.
\vskip-1mm
%%%%%%%%%%%%%%
%%%%%%

\item
%{Expanding universe}
Our last example is the motion in the expanding universe as seen by different observers \cite{Blau}.
Since Lukash metric is related with the open Friedmann
universe, we consider the flat $(k=0)$ FRW metric%
\begin{equation}
ds^{2}=-dt^{2}+a\left(t\right)^{2}\left(
d\rho^{2}+\rho^{2}d\Omega^{2}\right) .
\end{equation}%
With the conformal time%
\begin{equation}
dT =\frac{dt}{a(t)}
\end{equation}%
the above metric can be expressed using co-moving coordinates $(T,\rho,\theta,\phi)$, %
\begin{equation}
ds^{2}=a(T)^{2}\left(-dT^{2}+d\rho^{2}+\rho^{2}d\Omega^{2}\right) \,.
\label{comovingds}
\end{equation}

For the scale factor $a(t) =\sqrt{t}$, for example, one gets the radial trajectory $\theta,\phi =\const$,%
\begin{equation}
\rho(T) =\rho_{1}+\frac{\alpha}{2}\ln \left( 2T +\sqrt{4T^{2}+\alpha^{2}}\right),\quad \rho_{1} = \const
\end{equation}%
Thus for a co-moving observer staying in the co-moving coordinates $(T,\rho,\theta,\phi)$, the free particle can stay int rest~: $\rho(T) =\rho_{1} = \const
$ when $\alpha=0$.

Next, we consider global coordinates $\left(t,R,\theta ,\phi \right)$,
with global time $t$ and position %
\begin{equation}
R\left(t\right) = a\left(t\right)\rho=\sqrt{t}\,\rho\,.
\end{equation}%
Then the FRW metric  \eqref{comovingds} can be expressed as
\begin{equation}
ds^{2}=-\left(1-\frac{R^{2}}{4t^{2}}\right)dt^{2}-\frac{R}{t}%
dR\,dt+dR^{2}+R^{2}d\Omega^{2}
\end{equation}%
 with the radial trajectory becoming %
\begin{equation}
R\left( t\right) =\sqrt{t}\left\{\rho_{1}+\frac{\alpha }{2}\ln \left(4\sqrt{t}%
+4\sqrt{t+\alpha^{2}/16}\right) \right\}\,.
\end{equation}%
Thus in the global observer's coordinates $\left(t,R,\theta,\phi\right) $ the motion of the free particle is unbounded~: for a global observer in an expanding universe there is no bounded (localized) motion which, however, may exist for a co-moving observer.
%%%%%%
\eenu

We conclude that the  motion being bounded or unbounded may depend on the coordinates we choose.
%%%%%%%
\goodbreak

%%%%%%%%%%%%%%%%%%%%
\section{Conclusion}\label{Concl}
%%%%%%%%%%%%%%%%%%%%

\begin{figure}[ht]
\includegraphics[scale=.21]{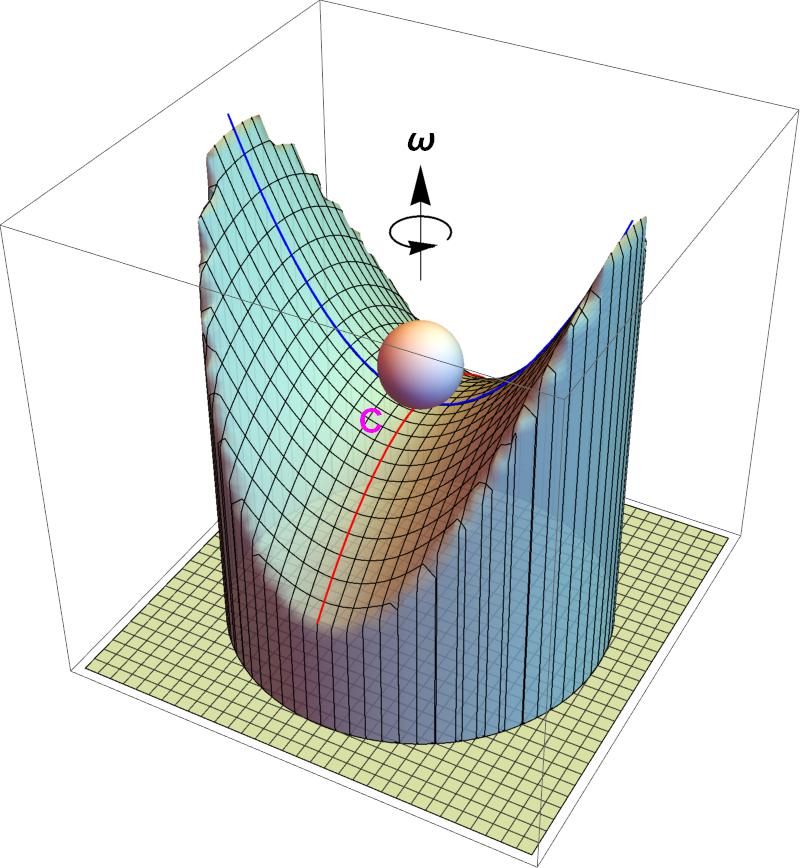}\\
\vskip-4mm
\caption{\textit{\small A ball put to the point $C$  will fall under  the slightest perturbation when the saddle is fixed, but its position will be stabilized when the saddle is rotated.
}
\label{saddlefig}
}
\end{figure}

Bialynicki-Birula and his collaborators
 argue that  gravitational waves emitted during the merger of a compact binary system may trap particles \cite{BB}.
Their statement is consistent with our previous study for a CPP wave \cite{POLPER,IonGW} for which we had found bounded geodesics. An analogy is provided by a rotating saddle \cite{PaulRMP,Kirillov,IonGW}, illustrated in FIG.\ref{saddlefig}.
%%%%%%%%%%%%%%%%%%
%%%%%%%%%%%%%%

Lukash gravitational waves were proposed to study anisotropic models \cite{LukashJETP75,SiklosAll,Collins%,EllisKing
}. Their profile in \eqref{BLukash} is reminiscent of but still different from (in fact more complicated) than that of CPP waves, \eqref{CPPprofile}.
They were studied in \cite{Lukash-I} along the lines set out by Siklos \cite{SiklosAll} in the parameter domain \eqref{BVII}, where they are of Bianchi type VII${}_h$. In this paper we consider instead what happens in the adjacent but different range, \eqref{boundcond} where problems are solved using different techniques, but, reassuringly, the results  agree  on the boundary,  \eqref{LLvalue}, which separates the parameter ranges.

%%%%%
One of our main results here is that the time redefinition \eqref{lnchange} relates CPP and Lukash waves schematically as,
\beq
\text{Lukash} \equiv  \text{CPP} \; + \; \text{linear force term}\,.
\label{LukCPP}
\eeq
%%%
Then a sequence of clever transformations  carries the time-dependent Sturm-Liouville problem to a system {with constant coefficients}, \eqref{abeqmot} - \eqref{TOmegas}\,,  we solve by chiral decomposition  \cite{Plyuchir,IonGW}.
Our results confirm  that particles can, in a suitable range of parameters, be trapped by Lukash waves, as
 shown in figs.\ref{boundfig}, \ref{Ccritfig}, \ref{C=kfig}, \ref{C>kfig}. Bounded geodesics arise when the wave is of Bianchi type VI.
The approximations used by \BB  lead to equations similar to our \eqref{abeqmot}.

%%%%%
 Our findings are exact in two respects~:
\begin{itemize}
\item
  We deal with exact plane gravitational waves, and do not use any weak field approximation.

\item We solve the equations of motion exactly.
\end{itemize}
%%%%%

Our time redefinition \eqref{lnchange} fits into the framework advocated  by Gibbons   \cite{GWG_Schwarzian}. Introducing new coordinates $(T,\bxi)$ by,
\beq
U=f(T),\qquad
\bX = \Big(\frac{df}{dT}\Big)^{1/2}\,\bxi
\label{txtauxi}
\eeq
extends the trick used in \cite{JunkerInomata,ZZH,Silagadze} in $d=1$ space dimension. Completing \eqref{lnchange} by \eqref{Vnushift} yields a conformal map between the perturbed CPP \eqref{Bmetric}-\eqref{Bprofile}-\eqref{CPPprofile} to Lukash,
$ds_{L}^{2} \leftrightarrow d{\Sigma}^{2}$, with conformal factor $\Omega^2(U)=e^{T}$.
This explains why \eqref{lnchange} works~:  conformally related space-times have, up to reparametrization, identical null geodesics.

As an additional bonus, the CPP $\leftrightarrow$ Lukash relation allows us to derive the symmetries of Lukash from those of CPP \cite{exactsol,Ehlers,POLPER}.
The additional 6th isometry \eqref{LUVboost} we call ``screw'' is reproduced  by following backwards our ``road map'' outlined. in sec. \ref{SLsol} and summarized, again schematically, by
\beq
(\bX, U, V) \quad \leftrightarrow \quad (\bxi, T, \nu) \quad \leftrightarrow\quad  (\binom{\alpha}{\beta}, T, \nu)\,.
\eeq

In our examples in sec.\ref{bubSec} the respective manifolds are conformally related, --- however  being bounded or not is  \emph{not invariant under a conformal redefinition of time}, \eqref{txtauxi} \footnote{This is analogous to the controversy which followed  the celebrated paper of Einstein and Rosen, as Iwo Bialynicki-Birula pointed out for us.}~: it is a coordinate dependent statement.

\begin{acknowledgments}\vskip-4mm
Correspondance and advice is acknowledged to Gary Gibbons and to Iwo Bialynicki-Birula.
M.E. is supported by the Bo\u{g}azi\c{c}i University Research Fund under grant number 21BP2 and
P-M.Z is supported by the National Natural Science Foundation of China (Grant No. 11975320).
\end{acknowledgments}
\goodbreak

%%%%%%%%%%%%%%%%%%%%%%%%%%%%%%%%%%%%%%%%%%%%%%%%%%%%%%%

\end{document}